\documentstyle[seceq,epsf]{ptptex}

\newcommand{\simgt}{\lower.5ex\hbox{$\; \buildrel > \over \sim \;$}}
\newcommand{\simlt}{\lower.5ex\hbox{$\; \buildrel < \over \sim \;$}}

\title{Strong~Gravitational~Lensing and Velocity~Function \\ 
  as~Tools~to~Probe Cosmological~Parameters: \\ Current~Constraints
  and Future~Predictions}

\author{Takahiro T. {\sc Nakamura}$^1$\footnote{%
    E-mail: nakamura@utaphp2.phys.s.u-tokyo.ac.jp}%
  and Yasushi {\sc Suto}$^{1,2}$\footnote{%
    E-mail: suto@phys.s.u-tokyo.ac.jp} }

\inst{$^1$Department of Physics, University of Tokyo, Tokyo, 113,
  Japan \\ $^2$Research Center for the Early Universe, University of
  Tokyo, Tokyo, 113, Japan }

\abst{ Constraints on cosmological models from strong gravitational
  lensing statistics are investigated. We pay particular attention to
  the role of the velocity function in the calculation of the lensing
  probability. The velocity function derived from the observed galaxy
  luminosity function, which is used in most previous work, is unable
  to predict the large separation lensing events. In this paper, we
  also use the Press-Schechter theory to construct a velocity function
  theoretically. Model predictions are compared with the observed
  velocity function and the HST snapshot survey. Comparison with the
  latter observation shows that the predictions based on the
  theoretical velocity function are consistent with the observed large
  separation events in COBE normalized low-density models, especially
  with a non-vanishing cosmological constant. Adopting the COBE
  normalization, however, we could find no model which simultaneously
  satisfies both the observed velocity function and the HST snapshot
  survey. We systematically investigate various uncertainties in the
  gravitational lensing statistics including finite core radius, the
  distance formula, magnification bias, and dust obscuration. The
  results are very sensitive to these effects as well as theoretical
  models for the velocity function, implying that current limits on
  the cosmological parameters should be interpreted with
  caution. Predictions for future surveys are also presented.  }

\begin{document}

\maketitle
\thispagestyle{empty}

\unitlength=0.01\textwidth
\begin{picture}(100,0)
\put(-5,70){\shortstack[l]{\sl
UTAP: University of Tokyo, Theoretical Astrophysics \\ \sl
RESCEU: Research Center for the Early Universe}
}
\put(80,70){\shortstack[l]{UTAP-236/96\\ RESCEU-23/96}}
\end{picture}

\vspace{-5mm}

\section{Introduction}\label{sec:intro}

To date, it is well known that the probability of gravitational
lensing (hereafter GL) of high redshift objects is sensitive to the
cosmological constant $\Lambda$ through the geometrical
effect\cite{AA,GPL,T90,FFK} and that extremely $\Lambda$-dominated
universes predict too many lensing events which is inconsistent with
the observed low frequency of lensed quasars in current
samples\cite{FT,FFKT,MR}. Kochanek\cite{K96}, for example, concludes
that $\lambda_0 := \Lambda/(3H_0^2) < 0.66$ at 95\% confidence.

GL limits on $\lambda_0$ should be contrasted to the other
cosmological tests which favor low density universes with
non-vanishing $\lambda_0$. The latter tests include the two point
angular correlation function of galaxies\cite{ESM}, the galaxy number
counts\cite{FTYY}, the cluster mass and peculiar velocity
functions\cite{BC92,UIS,BC93,BO} (See Refs.\cite{SU93,LLVW} for
reviews). Moreover, the recent measurements of the (local) Hubble
constant $H_0 =: 100h\,{\rm km\,s^{-1}} {\rm Mpc}^{-1}$ give somewhat
large values around $h\sim0.7$--$0.8$\cite{P94,F94,TSFR,WSLMB}. To
reconcile the age of the universe $\sim H_0^{-1}$ with that of the
oldest globular clusters\cite{C95} and of the young
galaxies\cite{D96}, it seems inevitable to introduce non-null
$\Lambda$ of $\lambda_0\simgt 0.8$\cite{NS}. Therefore it is
worthwhile reexamining the previous constraints on $\Lambda$ from GL
statistics by taking account of several uncertainties involved in the
calculation of GL probability.

In this paper, we pay particular attention to the role of the velocity
function (hereafter VF) of lensing objects. Previous work on GL
statistics usually uses VF determined from the observed galaxy
luminosity function. Moreover, it is often assumed that the comoving
number density of lensing objects is constant. However, the luminosity
function counts only luminous objects while invisible dark objects
might as well be responsible for GL. Bearing this in mind, we also use
the VF derived from the Press-Schechter theory\cite{PS} (hereafter PS)
with some specific cosmological models such as the cold dark matter
(CDM) and the primeval isocurvature baryon (PIB) models\cite{P87}.
The theoretical velocity function, by its construction, properly
counts all gravitationally bound dark halos as well as luminous
objects, and automatically takes account of the hierarchical evolution
of objects for various cosmological parameters\cite{LC93}. Narayan \&
White\cite{NW} first applied the PS theory to GL statistics, but only
in the standard CDM cosmology with no comparison with
observation. Kochanek\cite{K95} extended their calculation for various
cosmological parameters, and argued the constraints on the bias
parameter $b$ within the standard CDM model. However, we will see that
the PS theory in its original form is not a perfect one in deriving
VF, and adopt theoretical attempts towards more realistic VF.  We
compare these theoretical VF and their predictions on GL statistics
with the observed VF\cite{SH93} and the {\em HST snapshot
  survey}\cite{M93b} for COBE normalized CDM and PIB models.

We also make quantitative estimates of following uncertainties in GL
statistics: the effect of inhomogeneity of the universe in the
distance formula, the finite core size of lensing objects, the
obscuration of images by dusts in the lensing galaxies, and the
different definitions of the magnification bias. Although each topic
is already discussed separately\cite{FFKT,K96,ES,HK,FP,T93,ST}, we
take account of all the above effects consistently and systematically
for the first time.

This paper is organized as follows. In \S\ref{sec:gl}, we briefly
summarize the basic formulae of GL statistics. In \S\ref{sec:vf} we
present the VF's which are applied to GL statistics in the later
sections. In \S\ref{sec:obs}, we compare the theoretical VF and its GL
prediction with observation. In \S\ref{sec:unc}, we survey various
uncertainties described above and see how the results in
\S\ref{sec:obs} are affected by them. In \S\ref{sec:dss}, we give
predictions for future GL surveys. Finally in \S\ref{sec:sum}, we
summarize the main results of this paper. Our main results are
discussed after \S4 and are summarized in Table \ref{tab2}. Those who
are familiar with GL and the PS theory can skip \S2 and \S3 which are
written only for completeness and terminology.

Throughout the paper, we use the units $c=G=1$. The cosmological
parameters are defined as
\begin{equation}\label{1}
  \Omega := \frac{8\pi\bar\rho}{3H^2}, \quad \lambda :=
  \frac{\Lambda}{3 H^2}, \quad \Omega_0 :=
  \frac{8\pi\bar\rho_0}{3H_0^2}, \quad \lambda_0 :=
  \frac{\Lambda}{3H_0^2},
\end{equation}
\begin{equation}\label{2}
  H := \frac{\dot a}a = H_0 [\Omega_0 a^{-3} + (1-\Omega_0-\lambda_0)
  a^{-2} + \lambda_0 ]^{1/2},
\end{equation}
where $\bar\rho=\bar\rho_0/a^3=\bar\rho_0(1+z)^3$ is the mean density
of the universe at redshift $z$ and the subscripts 0 mean the present
epoch $z=0$.

This paper is largely based on Ref.\cite{N96} (unpublished) which
contains further detailed discussion.


\section{Brief summary of GL statistics}\label{sec:gl}

\subsection{Lens model and GL cross section}\label{sec:lensmod}

We assume that the lensing objects are spherical and isothermal in
describing GL statistics. The spherical lens model gives sufficiently
accurate description of the properties of lensed images such as the
image separation and the magnification, as compared with more
realistic elliptical lens models\cite{KK,BK,KB}. The isothermal
profile is characterized by the internal one-dimensional velocity
dispersion $v$\cite{BT}. Thereby we start from the lens
potential\cite{SEF} (twice the Newtonian potential integrated along
line-of-sight):
\begin{equation}\label{2.5}
  \hat\psi(\xi) = 4\pi v^2 (\xi^2 + \xi_{\rm c}^2)^{1/2},
\end{equation}
where $\xi$ is the impact parameter in the lens plane and the core
radius $\xi_{\rm c}$ eliminates the central singularity.\footnote{%
  The parameter $\xi_{\rm c}$ defined here approximately corresponds
  to twice the core radius of Ref.\cite{HK}.} The lens model with
$\xi_{\rm c}=0$ is the ``singular isothermal sphere'' (SIS). Due to
the spherical symmetry, the lens equation\cite{SEF} becomes a scalar
equation:
\begin{equation}\label{3}
  \eta = \frac{D_{\rm OS}}{D_{\rm OL}}\xi - D_{\rm LS}
  \hat\alpha(\xi),
\end{equation}
where $\eta$ is the source position in the source plane, $D_{\rm OL}$,
$D_{\rm OS}$ and $D_{\rm LS}$ are the angular diameter distances
between the observer, lens and source, and $\hat\alpha(\xi) :=
(d/d\xi) \hat\psi(\xi)$ is the deflection angle. The thin lens
approximation, which is implicit in Eq.(\ref{3}), can be
justified\cite{F93,SSE,PB} in the case of strong GL, because highly
virialized and compact galaxies or clusters are considered here as the
lensing objects. Defining
\begin{equation}\label{7n}
  \xi_* := 4\pi v^2 \frac{D_{\rm OL}D_{\rm LS}}{D_{\rm OS}}, \quad
  \eta_* := 4\pi v^2 D_{\rm LS}, \quad
\end{equation}
\begin{equation}
  x:= \xi/\xi_*\,, \quad y:=\eta/\eta_*\,, \quad x_{\rm c} := \xi_{\rm
    c}/\xi_*\,,
\end{equation}
one can rewrite Eq.(\ref{3}) as
\begin{equation}\label{6}
  y = x - x(x^2+x_{\rm c}^2)^{-1/2}.
\end{equation}
Given a source position $y$, the image position $x$ is obtained by
solving Eq.(\ref{6}). If $x_{\rm c}<1$ and $y<y_{\rm r} := (1-x_{\rm
  c}^{2/3}) ^{3/2} {\mit\Theta}(1-x_{\rm c})$, triple images form
[${\mit\Theta}(x)$ is the step function]. The brightness of an image
at $x$ is magnified by a factor\cite{SEF} $\mu_{\rm p}(x) := | (y/x)
(dy/dx) | ^{-1}$. When $x_{\rm c}\neq 0$, we solve Eq.(\ref{6})
numerically for the image positions $x_i(y)$ ($i=1,2,3$;
$x_1<x_2<x_3$) and calculate $\Delta\!x(y) := x_3-x_1$ (separation of
the outer two images), $\mu(y) := \sum_i \mu_{\rm p}(x_i)$ (total
magnification of all the images), and $r(y) : = \exp| \ln[\mu_{\rm
  p}(x_3) / \mu_{\rm p}(x_1)] |$ (brightness ratio [$>1$] of the outer
two images). As noted in Ref.\cite{HK}, it is safe to approximate
$\Delta\!x(y)\simeq \Delta\!x(0)=2x_{\rm t} : = 2(1-x_{\rm c}^2)^{1/2}
{\mit\Theta}(1-x_{\rm c})$. The separation angle $\theta$ in radian is
\begin{equation}\label{10}
  \theta = \frac{2\xi_*x_{\rm t}}{D_{\rm OL}}= 8\pi v^2 \frac{D_{\rm
      LS}}{D_{\rm OS}} (1-x_{\rm c}^2)^{1/2}.
\end{equation}
The GL cross section\cite{SEF} $\hat\sigma(Q)$ (defined as the area of
such a region in the source plane that a source is lensed with certain
properties $Q$ if it resides within the region) is generally written
as the integral in the $y$-plane:
\begin{equation}\label{8}
  \hat\sigma(Q) = 2\pi\eta_*^2\int_0^{\infty} S(y,Q)\,y\,dy\,.
\end{equation}
For example, if $Q$ is ``formation of triple images'', then
$S(y,Q)={\mit\Theta}(y_{\rm r}-y)$ and $\hat\sigma=\pi(\eta_*y_{\rm
  r})^2$; if $Q$ is ``image separation is larger than $\Delta\! x_*$''
{\em and} ``magnification is larger than $\mu_*$'' {\em and}
``brightness ratio is smaller than $r_*$'', then
\begin{equation}\label{11}
  \hat\sigma(\Delta\! x_*,\mu_*,r_*) = \eta_*^2 \,{\mit\Theta}(2x_{\rm
    t}-\Delta\! x_*) \sigma(\mu_*,r_*)\,,
\end{equation}
where
\begin{equation}\label{15}
  \sigma(\mu_*,r_*):=2\pi \int_0^{y_{\rm r}} dy\,y
  \,{\mit\Theta}\left(\mu(y)-\mu_* \right) \,{\mit\Theta}\left(
    r_*-r(y) \right)\,.
\end{equation}


\subsection{Distance formula}

The angular diameter distances ($D_{\rm OL},D_{\rm OS},D_{\rm LS}$)
are given by the solutions of the Dyer-Roeder
equation\cite{DR72,DR73,Z64}:
\begin{equation}\label{40n}
  \left[\frac{d^2}{dz^2} + \frac{3 + q(z)}{1+z}\frac{d}{dz} +
    \frac32\frac{\tilde\alpha\Omega(z)}{(1+z)^2} \right] D = 0\,,
\end{equation}
where $q:=\frac12\Omega-\lambda$ is the deceleration parameter, and
the ``smoothness parameter'' $\tilde\alpha$ ($0<\tilde\alpha<1$)
measures the degree of inhomogeneity of the universe (for derivation,
see Ref.\cite{SEF}). The last term of Eq.(\ref{40n}), the Ricci
focusing term, represents the matter density inside the beam of
photons. Denoting $D_{\tilde\alpha}(z_*,z)$ as the distance from $z_*$
to $z$ ($>z_*$), the initial conditions of the differential equation
(\ref{40n}) are
\begin{equation}
  D_{\tilde\alpha}(z_*,z_*)=0, \quad
  \frac{d}{dz}D_{\tilde\alpha}(z_*,z)\Bigm|_{z=z_*} =
  \frac1{(1+z_*)H(z_*)}.
\end{equation}
In particular, we write $D_{\tilde\alpha}(z) := D_{\tilde\alpha}
(0,z)$. The solution with $\tilde\alpha=1$ is the ``filled-beam
distance,'' and gives the standard angular diameter distance in an
exactly homogeneous FRW space-time\cite{W72}, while that with
$\tilde\alpha=0$ is the ``empty-beam distance.''  In the derivation of
Eq.(\ref{40n}), it is assumed that the focusing due to the tidal shear
(Weyl focusing) is negligible.  It is shown\cite{G67,FS,WS90} that
this assumption is plausible in most astronomical situations. In
appendix \ref{app:dist}, we present specific solutions of
Eq.(\ref{40n}).

\subsection{GL probability formula}

The probability that a source at redshift $z_{\rm s}$ is lensed with
properties $Q$ is generally written as\cite{SEF}
\begin{equation}\label{12}
  P(Q,z_{\rm s})= \frac1{D_1^2(z_{\rm s})}\int_0^{z_{\rm s}} dz
  \,\frac{(1+z)^2}{H(z)}D_1^2(z) \int d\chi\,
  \hat\sigma(Q,\chi,z,z_{\rm s}) N_{\chi}(\chi,z)
\end{equation}
where the integral variables $z$ and $\chi$ are, respectively, the
redshift and the parameter of the lensing objects, with the latter
completely characterizing their properties ($\chi$ is not necessarily
a one dimensional quantity), and $N_{\chi}(\chi,z)d\chi$ is the
``$\chi$-function'', i.e., the comoving number density of lensing
objects with the parameter $\chi\sim\chi+d\chi$ at $z$ (the reason why
the filled-beam distances are used in Eq.[\ref{12}] is discussed in
\S\ref{sec:dist}).

In the above lens model (Eq.[\ref{2.5}]), the lensing cross section is
characterized by a single parameter $v$ (assuming that $\xi_{\rm c}$
is empirically related to $v$ as Eq.[\ref{18}]). This is why VF plays
a fundamental role in the calculation of GL probability. Substituting
Eq.(\ref{11}) into Eq.(\ref{12}), the probability that a source at
$z_{\rm s}$ is multiple-imaged with image separation angle larger than
$\theta$ {\em and} total magnification larger than $\mu_*$ {\em and}
the brightness ratio smaller than $r_*$ is
\begin{equation}\label{16}
  P(\theta,\mu_*,r_*,z_{\rm s})= \frac{16\pi^2}{D_1^2(z_{\rm s})}
  \int_0^{z_{\rm s}} dz\, \frac{(1+z)^2}{H(z)} D_1^2(z)
  D_{\tilde\alpha}^2(z,z_{\rm s}) \int_{v_1}^{v_2} dv\, v^4
  N_v(v,z)\sigma(\mu_*,r_*)\,,
\end{equation}
The limits $v_1$ and $v_2$ ($>v_1$) of the $v$-integral are determined
by Eq.(\ref{10}) (see Eq.[\ref{b1}]). If $x_{\rm c}(v)$ does not
increase with $v$ (i.e., $\xi_{\rm c}$ increases less rapidly than
$\propto v^2$), $v_1$ is given by the sole positive solution of
Eq.(\ref{10}) for given $\theta$, and $v_2=\infty$ formally (Strictly
speaking, setting $v_2=\infty$ is inconsistent with the isothermal
lens model because the isothermal profile does not continue to
infinite radii. However, one can show that a truncation of isothermal
profile at the virial radius makes little difference). If $x_{\rm
  c}(v)$ increases with $v$, then $v_1$ and $v_2$ are two positive
solutions of Eq.(\ref{10}). Probability distribution of the image
separation angle is
\begin{eqnarray}
  P_{\theta}&&(\theta,\mu_*,r_*,z_{\rm s}) := -
  \frac{d}{d\theta}P(\theta,\mu_*,r_*,z_{\rm s}) \\ = &&
  \frac{16\pi^2}{D_1^2(z_{\rm s})} \int_0^{z_{\rm s}} dz
  \frac{(1+z)^2}{H(z)}D_1^2(z)D^2_{\tilde\alpha}(z,z_{\rm s})
  \left[\frac{dv}{d\theta} v^4 N_v(v,z)
    \sigma(\mu_*,r_*)\right]_{v_2}^{v_1} .
\label{26}
\end{eqnarray}
Assuming
\begin{equation}\label{18}
  \xi_{\rm c} = \xi_{\rm c*} v^p\,,
\end{equation}
$dv/d\theta$ in Eq.(\ref{26}) is calculated from Eq.(\ref{10}) as
\begin{equation}
  \frac{dv}{d\theta} = \frac{v}{\theta} \frac{1-x_{\rm c}^2(v)}{2 -
    px_{\rm c}^2(v)}\,.
\end{equation}
In \S\ref{sec:core}, we will see that observations suggest $p\sim
3$. In appendix \ref{app:v1v2}, we give analytic expressions for $v_1$
and $v_2$ in the cases $p=0,1,2,3$ and 4.


\subsection{Magnification bias}\label{sec:magb}

Because GL causes a magnification of images and bright QSO's are easy
to discover, lensed QSO's have relatively high probability of being
included in a QSO sample. This selection effect (magnification
bias\cite{T80}) is computed as follow. Let ${\mit\Phi}^{\rm Q}_L(L)dL$
be the luminosity function of sources, and let ${\mit\Phi}_{\rm Q}(L) :
= \int_L^{\infty} {\mit\Phi}^{\rm Q}_L(L')dL'$. When one searches for
lensed QSO's of the observed flux brighter than $S$, the GL
probability increases as\cite{TOG}:
\begin{equation}\label{20}
  P^{\rm B}_{\theta}(\theta,S,r_*,z_{\rm s}) := \frac1{{\mit\Phi}_{\rm
      Q}(L)} \int_1^{\infty} d\mu_*
  P_{\theta\mu_*}(\theta,\mu_*,r_*,z_{\rm s}) {\mit\Phi}_{\rm
    Q}(L\bar\mu/\mu_*)\,,
\end{equation}
where $P_{\theta\mu_*}:=-(d/d\mu_*)P_{\theta}$, $\bar\mu :=
[D_{\tilde\alpha}(z_{\rm s}) / D_1(z_{\rm s})]^2$, and the luminosity
\begin{equation}\label{45.5}
  L := 4\pi (1+z_{\rm s})^4D_1^2(z_{\rm s})(1+z_{\rm s})^{\gamma-1}\,S
\end{equation}
must be calculated from $S$ in the same way as in the determination of
${\mit\Phi}^{\rm Q}_L(L)$. The last factor $(1+z)^{\gamma-1}$ is the
$K$-correction\cite{P93}, which assumes that the energy spectrum of
QSO's is of the form $E \propto \nu^{-\gamma}$. We put the factor
$\bar\mu$ in Eq.(\ref{20}) because the ``unlensed'' sources are also
magnified (if $\tilde\alpha<1$) by that factor on average from the
flux conservation\cite{W76}. Substituting Eq.(\ref{26}) into
Eq.(\ref{20}), the biased probability $P^{\rm B}_{\theta}$ is
expressed as Eq.(\ref{26}) with $\sigma(\mu_*,r_*)$ replaced by
(cf. Eq.[\ref{15}]):
\begin{equation}\label{46}
  \sigma_{\rm B}(S,r_*) := \frac{2\pi}{{\mit\Phi}_{\rm Q}(L)}
  \int_0^{y_{\rm r}}dy\, y \,{\mit\Theta}\left( r_*-r(y) \right)
  \,{\mit\Phi}_{\rm Q}\left(\frac{\bar\mu L}{\mu(y)} \right)\,.
\end{equation}
Similarly, for a QSO of observed flux $S$, the GL probability is given
by Eq.(\ref{26}) with $\sigma(\mu_*,r_*)$ replaced by\footnote{%
  we use $(y_{\rm r}-y)^{(3 - \beta)/2}$ as an integral variable in
  Eqs.(\ref{45}) and (\ref{46}) in order to eliminate the divergence
  of the integrand on the radial caustics $y=y_{\rm r}$}
\begin{equation}\label{45}
  \sigma^{\rm B}_S(S,r_*) := \frac{2\pi}{{\mit\Phi}^{\rm Q}_L(L)}
  \int_0^{y_{\rm r}}dy\, y \,{\mit\Theta}\left( r_*-r(y) \right)
  \,{\mit\Phi}^{\rm Q}_L\left(\frac{\bar\mu L}{\mu(y)}\right)
  \frac{\bar\mu}{\mu(y)}
\end{equation}

The observed QSO luminosity function is fitted by the two power-law
model\cite{BSP}
\begin{equation}\label{47}
  {\mit\Phi}^{\rm Q}_L(L)dL = \phi_* [ (L/L_*)^{\alpha} +
  (L/L_*)^{\beta}]^{-1} (dL/L_*)
\end{equation}
with the luminosity evolution
\begin{equation}\label{48}
  L_*(z_{\rm s}) = L^*_0(1+z_{\rm s})^{k_{\rm L}}\,,
\end{equation}
where $\alpha=3.79$, $\beta=1.44$, $\phi_*=6.4\times 10^{-6} h^3 {\rm
  Mpc}^{-3}$, $k_{\rm L}=3.15$, and the absolute B-band magnitude
corresponding to $L^*_0$ is $M^*_0 = -20.91+5\log h$. These values are
obtained from $z_{\rm s}<2.2$ QSO sample, assuming $(\Omega_0,
\lambda_0) = (1,0)$ and $\gamma=0.5$ in Eq.(\ref{45.5}) (but the
values of $\alpha$ and $\beta$ do not depend on the assumed
cosmological models). Following Ref.\cite{WN}, we assume that
Eq.(\ref{48}) is valid up to $z_{\rm s}=3$ and that, for $z_{\rm
  s}>3$, $L_*(z_{\rm s}) = L^*_0 4^{k_{\rm L}}\,3.2^{(z_{\rm
    s}-3)/(\alpha-\beta)}$ and $\phi_*(z_{\rm s}) = \phi^*_0\,
3.2^{-(z_{\rm s}-3)(\alpha-1)/(\alpha-\beta)}$ with the same values of
the parameters. The behavior of this luminosity function at $z_{\rm
  s}>3$ is such that the bright end keeps constant while the faint end
slides down by a factor of 3.2 in every unit redshift. We also assume
that the QSO sample in Boyle et al.\cite{BSP} itself does not suffer
from the magnification bias, and neglect the possible magnification
due to microlensing by stars in the lensing galaxies\cite{BS}.


\section{Velocity function}\label{sec:vf}

\subsection{Schechter VF: observational viewpoint}\label{sec:schvf}

Standard calculation of GL probability in most of the previous work
uses VF from the observed galaxy luminosity function of Schechter form
at $z=0$:
\begin{equation}\label{30}
  N_L^{\rm Sch}(L)dL = \phi_* (L/L_*)^{-\alpha} e^{-L/L_*} dL/L_*\,.
\end{equation}
Autofib survey\cite{ECBHG} concludes that $\alpha=1.09$,
$\phi_*=0.026h^3 {\rm Mpc}^{-3}$ and the absolute magnitude
corresponding to $L_*$ is $M_*=-19.20+5\log h$ in B-band. Combining
with the Tully-Fisher or Faber-Jackson relation $L/L_* =
(v/v_*)^{\gamma}$, Eq.(\ref{30}) yields
\begin{equation}
  N^{\rm Sch}_v(v,0)dv = \sum_{i={\rm E,S0,S}}\phi_{*i}\gamma_i \left
    ( \frac{v}{v_{*i}} \right)^{-\gamma_i(\alpha-1)}\!\!  \exp\left[ -
    \left(\frac{v}{v_{*i}}\right)^{\gamma_i} \right] \frac{dv}{v}
\end{equation}
(hereafter SchVF). Therein constant comoving number density of lensing
objects, i.e., $N_v(v,z)=N_v(v,0)$, is often assumed. Using the B-band
Tully--Fisher and Faber--Jackson relations in Ref.\cite{SH93}, one
finds $(\gamma, v_*)=(2.9,126\,{\rm km\,s^{-1}})$ for S galaxies, and
$(\gamma, v_*)=(3.3, 175\,{\rm km\,s^{-1}})$ for E and S0 galaxies at
the above value of $L_*$. The morphological composition is $\phi_{*\rm
  E} + \phi_{*\rm S0} = 0.44\phi_*$ and $\phi_{*\rm S} =
0.56\phi_*$\cite{MHG}. We adopt the above values.

Several problems in applying SchVF to GL statistics include: (1) SchVF
counts only luminous objects though they are not the only lenses:
invisible dark haloes might as well be responsible for GL. (2) It is
possible that the assumption of no evolution makes the SchVF strongly
affected by the local and recent property of the universe around
us. In fact, Autofib survey\cite{ECBHG} found that the luminosity
function steepens with redshift up to $z=0.75$ at the faint end. Some
authors\cite{ST,M91,MK,RMTF} consider specific evolution models, but
their conclusions does not seem decisive in that the models include
some arbitrary free parameters. [Inclusion of dark objects (1) and the
evolution effect (2) into GL statistics would, in principle, lead to
tighter limits on $\Lambda$.] (3) The Tully-Fisher and Faber-Jackson
relations may not be universal in particular at high $z$. (4) SchVF
has an uncertainty in the velocity dispersion of early-type galaxies
by a factor $1.5$\cite{G77}, which changes the GL probability by
$(1.5)^2$. [See Ref.\cite{K96} for a detailed study against these
arguments.] (5) Clusters of galaxies, which are not counted in SchVF,
can also contribute to strong lensing of QSO's\cite{FLO,T96}. Because
of these uncertainties, we would like to introduce another approach
below to the computation of VF.


\subsection{Press-Schechter VF: theoretical viewpoint}\label{sec:psvf}

Narayan \& White\cite{NW} and Kochanek\cite{K95} applied the PS mass
function (see, e.g., Ref.\cite{PAD} for introduction) to compute VF:
\begin{equation}\label{31}
  N_M^{\rm PS}(M,z)dM = \sqrt{\frac2{\pi}} \frac{\bar\rho_0}{M}
  \frac{\delta_{\rm c0}(z)}{\sigma(R)} \left| \frac{d\ln\sigma}{d\ln
      M} \right| \exp \left[-\frac{\delta^2_{\rm
        c0}(z)}{2\sigma^2(R)}\right]\frac{dM}M\,,
\end{equation}
where
\begin{equation}\label{32}
  \sigma(R) = \frac1{2\pi^2}\int_0^{\infty} dk\, k^2 P(k)W^2(kR)
\end{equation}
is the rms of linear density fluctuation today on the comoving scale
$R=[2M/(\Omega_0H_0^2)]^{1/3}$ [$P(k)$ and $W(kR)$ are the power
spectrum at $z=0$ and the $k$-space window function] and $\delta_{\rm
  c0}(z)$ is the critical density contrast extrapolated linearly to
today in order to virialize by $z$ (Eq.[\ref{c29.5}]). Assuming the
isothermality of the lensing objects again, the one-dimensional
velocity dispersion is related to the mass through
\begin{equation}\label{33} 
  v = \left[\frac{M}{2r_{\rm v}}\right]^{1/2}
  =\frac12[\Omega_0\vartheta_{\rm v0}^{1/3}(z)]^{1/2} H_0 R
\end{equation}
where $r_{\rm v}$ is the virial radius of the object and
$\vartheta_{\rm v0}(z):=(R/r_{\rm v})^3$ is the overdensity of an
object at $z=0$ which virialized at $z$
(Eq.[\ref{c29.4}]). Calculation of $\delta_{\rm c0}$ and
$\vartheta_{\rm v0}$ in $\lambda=0$ universe is found in
Ref.\cite{LC93}. We give analytic formulae for them in
$\Omega+\lambda=1$ universe in appendix \ref{app:sph}, using the
spherical collapse model. We normalize $\sigma(R)$ so that $\sigma
(8h^{-1}{\rm Mpc}) = b^{-1}$ ($b$ is the bias parameter) and use the
top-hat window function $W(x)=3(\sin x - x\cos x)/x^3$. From
Eqs.(\ref{31}) and (\ref{33}), VF is constructed theoretically
as\cite{CK,WF} (hereafter PSVF):
\begin{equation}\label{34}
  N_v^{\rm PS}(v,z)dv = \frac3{(2\pi)^{3/2}}
  \frac1{R^3}\frac{\delta_{\rm c0}(z)}{\sigma(R)}
  \left|\frac{d\ln\sigma}{d\ln R}\right| \exp \left[ -
    \frac{\delta_{\rm c0}^2(z)}{2\sigma^2(R)} \right]\frac{dv}{v}
\end{equation}
with the $v$-$R$ relation in Eq.(\ref{33}). PSVF, by its construction,
properly counts all gravitationally bound objects (including the
clusters of galaxies) as well as luminous objects, and automatically
takes account of the hierarchical evolution of objects in the universe
for various cosmological parameters\cite{LC93}. The theoretical
foundation of the PS mass function is now more secure\cite{BCEK} and
it is also supported by cosmological $N$-body
simulations\cite{EFWD,LC94}.

PSVF strongly increases with $\Omega_0$ and weakly dependent on
$\lambda_0$. The former is because the comoving scale $R$ is smaller
for larger $\Omega_0$ for fixed $v$ (Eq.[\ref{33}]): since the density
fluctuation is large on small scales [$\sigma(R)$ is large], larger
$\Omega_0$ universes have more objects at fixed $v$. The latter is
because $\delta_{\rm c0}$ and $\vartheta_{\rm v0}$ do not so much
depend on $\lambda_0$ at low redshifts. At higher redshifts, however,
$\delta_{\rm c0}$ and $\vartheta_{\rm v0}$ are more dependent on
$\lambda_0$ and so is PSVF. The bias parameter $b$ affects PSVF most
strongly: since $\sigma(R)\propto b^{-1}$, PSVF behaves like $\propto
b\exp(-b^2)$. The dependence on $h$ and $\Omega_{\rm B}$ is very weak
because they enter only through $\mit\Gamma$ (see Eq.[\ref{36n}])


\subsection{Formation rate in theoretical VF}\label{sec:ksvf}

PSVF is derived on the assumption that all objects have just
virialized at the moment when one evaluates VF\cite{S94,VL} (see the
definition of $\vartheta_{\rm v0}$). In the application to GL
statistics here, this means that the deflection of light and the
formation (virialization) of the deflector occurs simultaneously. In
reality, however, a virialized object survives for some time until it
is merged into more massive objects\cite{LC93,KS}, and hence there
must be an accumulation of objects in a specific velocity range which
formed at some earlier epochs. Therefore, we must take account of the
formation epoch distribution of objects in the universe to avoid the
above assumption.

Kitayama \& Suto\cite{KS} proposed a practical prescription in
obtaining the formation epoch distribution from the random-walk
method\cite{BCEK}, and calculated theoretically the comoving number
density $F(M,z_{\rm f},z)dM dz_{\rm f}$ of those objects which formed
at $z_{\rm f}\sim z_{\rm f}+dz_{\rm f}$ with mass $M\sim M+dM$ {\em
  and} survive without destructed (absorbed into larger hierarchies)
until $z$. Integrating over $z_{\rm f}$, more realistic VF may be
written as (hereafter KSVF):
\begin{equation}\label{40}
  N_v^{\rm KS}(v,z)dv = dv \int_z^{\infty} dz_{\rm f} F(M,z_{\rm f},z)
  \frac{dM}{dv}\,.
\end{equation}
The explicit form of $F$ is found in their Eq.(23).  Compared with
PSVF, KSVF is shifted to the direction of larger velocity relative to
PSVF, especially on small scales (see Figs.\ref{fig1} and \ref{fig2}).
This is due to the accumulation of objects, as mentioned above:
because smaller objects (at $z=0$) are likely to have formed earlier
in the bottom up scenario, and because older objects have higher
velocities (see Eq.[\ref{33}]), the velocity of small objects becomes
larger when we take the formation history into account.  Eq.(\ref{40})
refines the original PSVF in the sense that the formation epoch is
taken into account, though not fully satisfactorily\cite{KS}.


\section{Comparison with observation}\label{sec:obs}

\subsection{Comparison with observed VF}\label{sec:vfobs}

In this section, we compare theoretical VF in \S\ref{sec:vf} with the
observed one at $z=0$\cite{SH93}.
\begin{table}[htb]
\caption{Model parameters}\label{tab1}
\bigskip
\begin{tabular}{cccccccc}
\hline
Model & $P(k)$ & $\Omega_0$ & $\lambda_0$ & $h$ & $b^{-1}$ &
$\Omega_{\rm B} h^2$ \\ \hline  
SC & CDM & 1   & 0   & 0.5 & 1.3  & 0.016 \\
OC & CDM & 0.3 & 0   & 0.7 & 0.52 & 0.016 \\
LC & CDM & 0.2 & 0.8 & 0.8 & 1.0  & 0.016 \\
LP & PIB & 0.2 & 0.8 & 0.8 & 1.0  & 0.128 \\
\hline
\end{tabular}
\end{table}
We consider four specific cosmological models listed in Table
\ref{tab1}. The first three models (SC, OC and LC) use the scale
invariant ($n=1$) power spectrum of the cold dark matter (CDM) model
in Ref.\cite{BBKS}, and the last model (LP) uses that of the primeval
isocurvature baryon\cite{P87} (PIB) model in Ref.\cite{HBS} with the
slope $n=-1.15$ at the large wavenumber. The Hubble constant $h$ in
each model is chosen so that the age of the universe be longer than 13
Gyr\cite{C95} for given ($\Omega_0,\lambda_0$).  The bias parameters
$b$ adopts the COBE DMR 2yr data normalization\cite{S95}, and the
baryon density parameter $\Omega_{\rm B}$ in the CDM models is from
the big-bang nucleosynthesis constraint\cite{CST}. The $\mit\Gamma$
parameter in the CDM model is empirically fitted as\cite{S95}:
\begin{equation}\label{36n}
  {\mit\Gamma} = \Omega_0 h (T_0/2.7{\rm K})^{-2}\exp[-\Omega_{\rm
    B}(1+\sqrt{2h}/\Omega_0)]
\end{equation}
where $T_0\simeq2.726$K is the temperature of the cosmic microwave
background.

\begin{figure}[htb]
\epsfxsize=8cm\centerline{\epsfbox{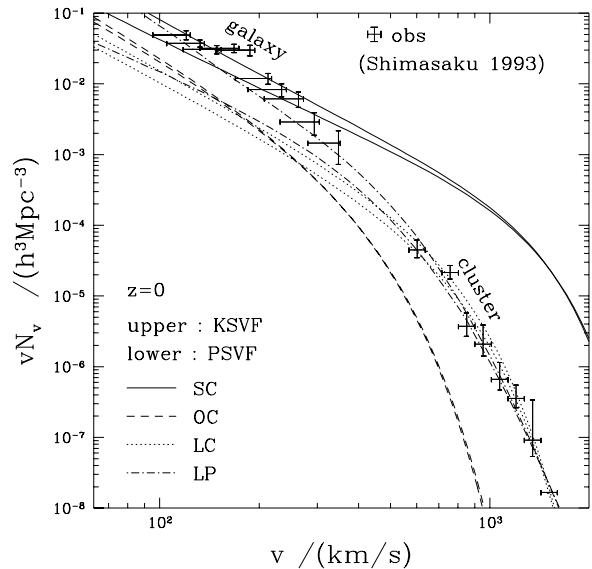}}
\caption{PSVF (Eq.[\protect\ref{34}]) and KSVF (Eq.[\protect\ref{40}])
  in the cosmological models listed in Table
  \protect\ref{tab1}. Crosses show the observed VF in
  Ref.\protect\cite{SH93}.}\label{fig1}
\end{figure}

In Fig.\ref{fig1}, we plot the theoretical VF's (PSVF and KSVF) for
these models at $z=0$. In $\Omega+\lambda=1$ case, we used the fitting
formulae for $\delta_{\rm c0}$ and $\vartheta_{\rm v0}$ in
Eqs.(\ref{c19}) and (\ref{c32}). Plotted in crosses are the observed
VF compiled by Shimasaku\cite{SH93}. He multiplied the observed
central velocity dispersion of stars in E and S0 galaxies by a
$1.5^{1/2}$ factor to take account of the difference between the
distributions of luminous matter and isothermal dark halos.  Since the
error bars in the raw data represent the statistical ones only, we
lengthened the error bars of galaxy VF by the $1.5^{1/2}$ factor into
the smaller velocity direction. However, it should be cautioned that
this is an over-modification, especially at the low velocity end,
because S galaxies do not have the factor. The cluster VF is
translated from the observed temperature function of Ref.\cite{ESFA}
through the isothermal $\beta$-model assuming $\beta=1$.

At least the VF observation should give a lower bound of the
theoretical VF because the observed VF counts only luminous
objects. In Fig.\ref{fig1}, model SC fits the galaxy VF fairly well,
but predicts too many large scale objects. Model OC fails on all
scales due to the small value of the normalization $b^{-1}$ from the
COBE data. Models LP and LC are very good on cluster scale but
significantly underpredict galactic size objects, except for the KSVF
in model LP.  Shimasaku\cite{SH93} also compared PSVF with the
observation, and concluded that no COBE normalized models are
consistent with VF observation within PSVF. However, as noted in
\S\ref{sec:ksvf}, PSVF has a flaw: PSVF neglects the accumulation of
objects which formed earlier. Although KSVF and PSVF are not so
different within the CDM models, they differ significantly in model
LP. This originates from the fact that the PIB models have more powers
on small scales than CDM models have: since smaller objects survive
longer, the accumulation in PIB models is more remarkable than that in
CDM models, especially on small scales.

Nearly on all scales, KSVF in model LP seems consistent with
observation if we allow the error bars to exclude the $1.5^{1/2}$
factor. However, it should be borne in mind that there underlies a lot
of uncertainties both in theory and observation of VF.  On the
observational side: (1) The $1.5^{1/2}$ factor is quite generous and
may represent the error considerably larger than the actual one. (2)
Although the galaxies in the Virgo cluster is excluded in the
Shimasaku\cite{SH93} data, it is possible that the galaxies in small
groups are counted as individual objects in that data. From the
construction of theoretical VF's, a group of galaxies should be
counted as one object and the galaxies inside should be neglected. (3)
The assumption of $\beta=1$ for cluster VF may not be justified
because of the $\beta$-discrepancy problem\cite{BL}.  (4) The observed
VF may not represent the cosmic mean. On the theoretical side: (5)
KSVF has some problems in counting argument\cite{KS}. (6) We have not
taken any dissipative processes into account in the dynamics of
virialization, both in PSVF and KSVF. That is, we have related mass
with velocity (Eq.[\ref{33}]) only from a virial analysis and
neglected a possible dissipation of energy from the collapsing object.

Here let us discuss the last point (6) at some length. If one takes
these dissipative processes into account, the velocity may increase
because of the contraction to smaller radii. Based on the arguments of
cooling and dynamical time scales\cite{RO,WR}, these effects are
expected to be unimportant on cluster scales. On galactic scales where
the cooling is likely to be effective, theoretical VF will shift to
the right and should become steeper as compared with the
dissipationless case. This line of arguments might make the model LC
consistent with the observed VF. We do not estimate here
quantitatively to what extent the dissipation modifies the theoretical
VF, but only note that the degree of dissipative effects in the galaxy
formation can be measured by comparing the observed VF with the
theoretically best VF without dissipation.

In considering GL statistics, it is desirable to use VF's which are
theoretically well-motivated and consistent with the observed VF. In
particular, the evolution of the VF's is important in the calculation
below. If one adopts the observed VF, however, one has to assume an
ad-hoc number evolution.  Given those uncertainties and problems, we
have decided to apply theoretical and observed VF's in computing the
GL probabilities on an equal basis.  Further comparison between the
theory and observation of VF should wait for future VF observation at
higher redshifts.

\begin{figure}[htb]
\epsfxsize=12cm\centerline{\epsfbox{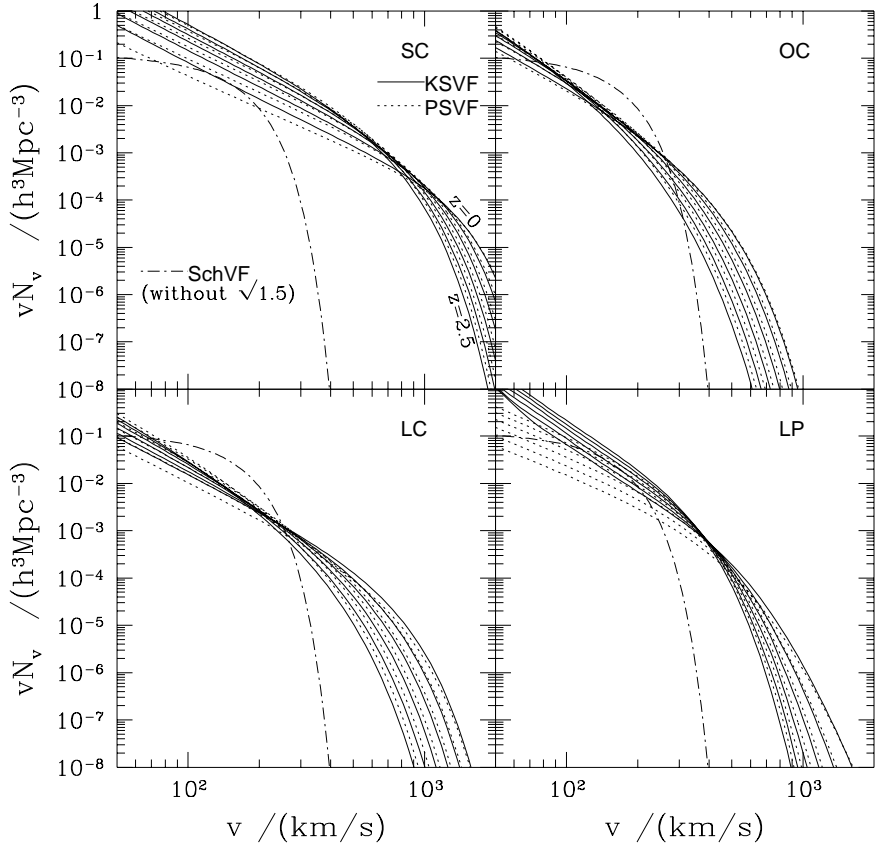}}
\caption{Evolution of KSVF (solid) and PSVF (dotted) for the
  cosmological models listed in Table \protect\ref{tab1}. SchVF
  without $1.5^{1/2}$ is also shown in dot-dashed curve for reference.
  From upper to lower curves in high velocity end, $z=0$, 0.5, 1, 1.5,
  2, 2.5.}
\label{fig2}
\end{figure}

In Fig.\ref{fig2} we plot the number evolution of PSVF and KSVF for
the cosmological models in Table \ref{tab1}. The decrease of small
scale objects is slower in KSVF than that in PSVF, especially in
models LC and LP.  This is because the accumulation of small objects
becomes less significant for higher redshifts in KSVF. The two VF's
become almost identical at the redshift of $\sim 4$. Recently, Mo \&
Fukugita\cite{MF} compared PSVF with the abundance of giant galaxies
at $z=3$, and conclude that the spatially flat CDM models with
$\lambda_0\sim 0.7$ are favorable. However, their conclusion seems
premature in that the model is already inconsistent with observation
at $z=0$ (Fig.\ref{fig1}).


\subsection{Comparison with the HST snapshot survey}\label{sec:hst}

Let us now calculate the GL probability in the $\xi_{\rm c}=0$ (SIS
lens model) case, and compare the result with the observed GL
frequency in the {\em HST snapshot
  survey}\cite{B92,M92b,M93a,M93b}. Various uncertainties including
the non-zero core radius are examined in the next section.  Although
there exist different surveys in optical band\cite{CMF,SUR93,JJPS}, we
do not mix these data to obtain the larger number of samples\cite{K96}
because we are mainly concerned with how the results are affected by
various uncertainties described in \S\ref{sec:vfobs} and
\S\ref{sec:unc}.

Using the redshifts and V-band magnitudes of all the 502 QSO's in the
HST survey, we calculate the expected number of lensed QSO's in the
survey for various cosmological parameters as
\begin{equation}\label{36}
  n_{\theta}^{\rm exp}(\theta) := \sum_{i=1}^{502}P^{\rm
    B}_{\theta}(\theta,S_i,r_{*i},z_i)
\end{equation}
with Eqs.(\ref{20}) and (\ref{45}), where $S_i$ and $z_i$ are the
observed V-band flux and redshift of the $i$-th QSO in the survey.
The HST survey can detect multiple images if $\theta>0.1''$ and if the
brightness ratio is smaller than\cite{MR}
\begin{equation}\label{37}
  r_{*i} = \min[19(\theta/{\rm arcsec})^{0.85}, S_i/S_{\rm lim},
  40]\,,
\end{equation}
where $S_{\rm lim}$ is the faintest flux that the HST can detect (the
corresponding apparent magnitude is $m_{\rm lim}=22$ in V-band). We
assume $B-V=0.2$\cite{B92} when the V-band magnitudes are inserted
into the B-band QSO luminosity function Eq.(\ref{47}). The
luminosities $L_i$ of QSO's are calculated from $S_i$ by
Eq.(\ref{45.5}) with $(\Omega_0,\lambda_0)=(1,0)$ and
$\gamma=0.5$\cite{BSP,SG}. We use the standard angular diameter
distance ($\tilde\alpha=1$; filled beam) for
$D_{\tilde\alpha}(z,z_{\rm s})$ in Eq.(\ref{26}).

\begin{figure}[htb]
\epsfxsize=8cm\centerline{\epsfbox{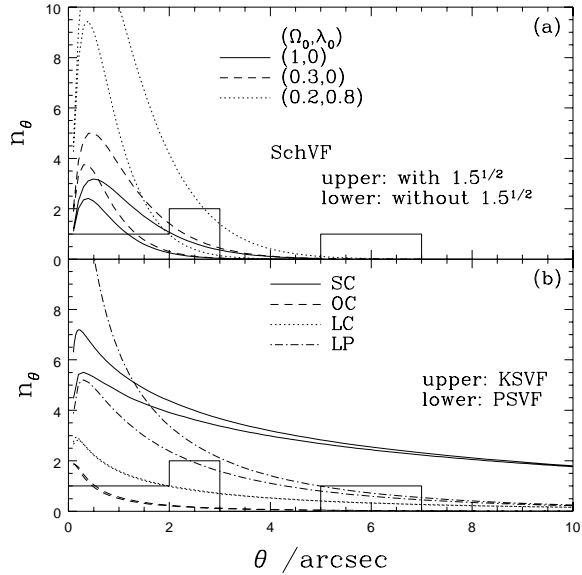}}
\caption{Image separation distribution of the expected number of
  lensed QSO's in the HST survey. (a): predictions from SchVF. (b):
  predictions from PSVF and KSVF. Model parameters in (b) are listed
  in Table \protect\ref{tab1}. Histogram shows the observed
  distribution. SIS lens model ($\xi_{\rm c}=0$) and the filled-beam
  distance ($\tilde\alpha=1$) are used.}
\label{fig3}
\end{figure}

Fig.\ref{fig3}a shows the prediction based on SchVF, while in
Fig.\ref{fig3}b we plot the PSVF and KSVF predictions for the model
parameters listed in Table\ref{tab1}. In models OC and LC, the KSVF
and PSVF predictions are almost indistinguishable (see
Fig.\ref{fig1}). Shown in the histogram is the distribution of the
observed 6 GL candidates in the HST survey. These are 1208+1011,
1413+117, 1115+080, 0142$-$100, 0957+561, 1120+0154, and the image
separation angles are $0.47''$, $1.22''$, $2.0''$, $2.2''$, $5.7''$
and $6.6''$, respectively. Despite the high resolution imaging of the
HST Planetary Camera, 1208+1011 is the only sub-arcsecond GL candidate
found by the survey\cite{M92a,MSVPH}.

Note that the SchVF can never predict the large separation lensing
events such as 0957+561 and 1120+0154, even if the $1.5^{1/2}$ factor
is included. Usually these large separation events are neglected as
statistical flukes in the literatures of GL statistics based on SchVF.
Maoz \& Rix\cite{MR}, for example, neglects 0957+561 because it is not
the lensing by a single galaxy but the lensing galaxy may be embedded
in a cluster\cite{YGKOW}. They also neglect 1120+0154 because the
images might be a physical pair (however, see
Refs.\cite{MD,MOM}). However, we believe that it is important to
consider these large separation lensing events when the PSVF or KSVF
predictions are compared with observation, because the theoretical
VF's in principle count all the gravitationally bound dark objects as
well as luminous objects. The apparent absence of any obvious lensing
object in 1120+0154 may be due to such non-luminous massive objects.

It is also interesting to note that the predictions from observed VF
and theoretical VF have opposite dependence on $\Omega_0$. This is
because theoretical VF naturally decreases with $\Omega_0$ (low
density universe has fewer lenses). On the contrary, since SchVF does
not at all depend on the cosmological parameters but is fixed by
observation, the $\Omega_0$ and $\lambda_0$ dependence of SchVF
prediction enters only through the cosmological geometry (distance
formulae and the Hubble constant). In the case of PSVF and KSVF
predictions, the $\lambda_0$ dependence comes mainly from the COBE
normalized bias parameter $b$.

Fig.\ref{fig3}a shows that the SchVF prediction apparently prefers
$\Lambda=0$ models to highly $\Lambda$-dominated
models\cite{FT,FFKT,MR,K96}. In Fig.\ref{fig3}b, model SC predicts too
many lensing events in all separation range\cite{K95}. Again model OC
is not viable: the COBE normalized density fluctuation amplitude
$b^{-1}$ in this model is too small to accommodate sufficient lensing
objects. The most likely values of $\Omega_0$ in open universes should
lie between 0.3 and 1. Model LC seems relatively good, while model LP
is good at large separations but too much overpredicts the small
separation events. We tabulate in Table \ref{tab2} the expected number
$n_{\rm exp} := \int n_{\theta}^{\rm exp}(\theta) d\theta$ of lensed
sources for these models, and their $\chi^2$ significance
level\cite{PTVF}
\begin{equation}\label{35}
  Q := \frac{ {\mit\Gamma}(n_{\rm bin}/2, \chi^2/2) } {
    {\mit\Gamma}(n_{\rm bin}/2) }, \quad \chi^2 := \sum_{i=1}^{n_{\rm
      bin}} \frac{(n^{\rm obs}_i - n^{\rm exp}_i)^2}{n^{\rm exp}_i}\,,
\end{equation}
where $n^{\rm obs}_i$ and $n^{\rm exp}_i$ are the observed and
expected number of lensed sources in the $i$-th bin. We set the bin
width to be $\Delta\!\theta_{\rm bin}=1''$ as shown in Fig.\ref{fig3},
and calculate $n_{\rm exp}$ and $Q$ within the separation range
$0''<\theta<4''$ for SchVF predictions, and within $0''<\theta<10''$
for PSVF and KSVF predictions (see the first row of the table denoted
as ``standard'').

We note that, if the $1.5^{1/2}$ factor is excluded, the $\Lambda=0$
models in the SchVF prediction have very low significance level,
because these models do not predict sufficient number of lensed QSO's
around $\theta\sim2''$ (1115+080 and 0142$-$100). If the $1.5^{1/2}$
factor is included, the values of $Q$ in $\Lambda=0$ models are larger
than those in $\Lambda\neq 0$ models {\em only relatively}: no models
in the SchVF prediction can fit the observed distribution
satisfactorily. The maximum likelihood analysis\cite{K93} is unable to
show this fact, where only the relative values of the likelihood among
the models are calculated. This is why we do not use the likelihood
analysis for the cosmological parameters.\footnote{%
  We do use the Kolmogorov--Smirnov test here because we want to
  compare the amplitude of the observed event rate with that of the
  theoretical predictions. This is also why we use the $\chi^2$ test
  although the binning is very crude.} We caution, however, that the
probabilities in the table are not statistically so meaningful because
of the small number of lensed candidates. More reliable tests will be
possible after the next-generation GL samples become available (see
\S\ref{sec:dss}).

\begin{table}[htb]
\caption{The expected number $n_{\rm exp}$ of lensed QSO's in the HST
  snapshot survey and the $\chi^2$ significance level $Q$
  (Eq.[\protect\ref{35}]) of the models (see Table
  \protect\ref{tab1}). (a,b): SchVF predictions within the separation
  range $0'' < \theta < 4''$; (c): PSVF predictions within $0'' <
  \theta < 10''$; (d): KSVF predictions within $0'' < \theta < 10''$;
  are compared with the observation.}
\label{tab2}

\bigskip

\begin{tabular}{|l|cc|cc|cc|}
\hline
(a)\hfill VF & 
\multicolumn{6}{c|}{SchVF without $1.5^{1/2}$} \\ \hline
\hfill ($\Omega_0,\lambda_0$) &
\multicolumn{2}{c|}{(1,0)} & \multicolumn{2}{c|}{(0.3,0)} &
\multicolumn{2}{c|}{(0.2,0.8)} \\ \hline
\hfill $0''<\theta<4''$ ($n_{\rm obs}=4$) & $n_{\rm exp}$ & $Q$ &
$n_{\rm exp}$ & $Q$ & $n_{\rm exp}$ & $Q$ \\ \noalign{\hrule height0.8pt}
Standard (\S\ref{sec:hst})&
2.5 & $<0.1\%$ & 3.7 & $<0.1\%$ & 10 & 2.1\% \\ \hline
Distance formula$^{\rm a}$ (\S\ref{sec:dist})& 
1.4 & $<0.1\%$ & 2.9 & $<0.1\%$ & 7.2 & 0.5\% \\ \hline
Core radius$^{\rm b}$ (\S\ref{sec:core})& 
2.3 & $<0.1\%$ & 3.4 & $<0.1\%$ & 9.4 & 1.2\% \\ \hline
Magnification bias$^{\rm c}$ (\S\ref{sec:brfn})& 
1.7 & $<0.1\%$ & 1.9 & $<0.1\%$ & 5.3 & 0.5\% \\ \hline
Dust obscuration$^{\rm d}$ (\S\ref{sec:dust})& 
1.8 & $<0.1\%$ & 2.7 & $<0.1\%$ & 7.8 & 4.2\% \\ \hline
\end{tabular}

\bigskip

\begin{tabular}{|l|cc|cc|cc|}
\hline
(b)\hfill VF & 
\multicolumn{6}{c|}{SchVF with $1.5^{1/2}$} \\ \hline
\hfill ($\Omega_0,\lambda_0$) &
\multicolumn{2}{c|}{(1,0)} & \multicolumn{2}{c|}{(0.3,0)} &
\multicolumn{2}{c|}{(0.2,0.8)} \\ \hline
\hfill $0''<\theta<4''$ ($n_{\rm obs}=4$) & $n_{\rm exp}$ & $Q$ &
$n_{\rm exp}$ & $Q$ & $n_{\rm exp}$ & $Q$ \\ \noalign{\hrule height0.8pt}
Standard (\S\ref{sec:hst})&
5.1 & 38\% & 7.5 & 30\% & 20 & 0.7\% \\ \hline
Distance formula$^{\rm a}$ (\S\ref{sec:dist})& 
2.8 & 3.7\% & 5.9 & 25\% &15 & 5.2\% \\ \hline
Core radius$^{\rm b}$ (\S\ref{sec:core})& 
4.4 & 26\% & 6.5 & 26\% & 18 & 1.5\% \\ \hline
Magnification bias$^{\rm c}$ (\S\ref{sec:brfn})& 
2.6 & 9.1\% & 3.8 & 19\% & 11 & 19\% \\ \hline
Dust obscuration$^{\rm d}$ (\S\ref{sec:dust})& 
4.1 & 45\% & 6.0 & 43\% & 17 & 2.7\% \\ \hline
\end{tabular}

\bigskip

\begin{tabular}{|l|cc|cc|cc|cc|}
\hline
(c)\hfill VF & \multicolumn{8}{c|}{PSVF} \\ \hline
\hfill Model & \multicolumn{2}{c|}{SC} & 
\multicolumn{2}{c|}{OC} & \multicolumn{2}{c|}{LC} &
\multicolumn{2}{c|}{LP} \\ \hline
\hfill $0''<\theta<10''$ ($n_{\rm obs}=6$) & $n_{\rm exp}$ & $Q$ &
$n_{\rm exp}$ & $Q$ & $n_{\rm exp}$ & $Q$ & $n_{\rm exp}$ & $Q$ \\
\noalign{\hrule height0.8pt} 
Standard (\S\ref{sec:hst})& 
29 & 3.2\% & 1.7 & $<0.1\%$ & 6.1 & 79\% & 16 & 45\% \\ \hline
Distance formula$^{\rm a}$ (\S\ref{sec:dist})& 
17 & 55\% & 1.3 & $<0.1\%$ & 4.6 & 48\% & 9.3 & 81\% \\ \hline
Core radius$^{\rm b}$ (\S\ref{sec:core})& 
18 & 44\% & 1.6 & $<0.1\%$ & 5.4 & 55\% & 11 & 73\% \\ \hline
Magnification bias$^{\rm c}$ (\S\ref{sec:brfn})& 
23 & 17\% & 1.4 & $<0.1\%$ & 5.1 & 60\% & 11 & 79\% \\ \hline
Dust obscuration$^{\rm d}$ (\S\ref{sec:dust})& 
27 & 6.5\% & 1.3 & $<0.1\%$ & 5.7 & 79\% & 12 & 79\% \\ \hline
\end{tabular}

\bigskip

\begin{tabular}{|l|cc|cc|cc|cc|}
  \hline (d)\hfill VF & \multicolumn{8}{c|}{KSVF} \\ \hline \hfill
  Model & \multicolumn{2}{c|}{SC} & \multicolumn{2}{c|}{OC} &
  \multicolumn{2}{c|}{LC} & \multicolumn{2}{c|}{LP} \\ \hline \hfill
  $0''<\theta<10''$ ($n_{\rm obs}=6$) & $n_{\rm exp}$ & $Q$ & $n_{\rm
    exp}$ & $Q$ & $n_{\rm exp}$ & $Q$ & $n_{\rm exp}$ & $Q$ \\ 
  \noalign{\hrule height0.8pt} Standard (\S\ref{sec:hst})& 32 & 1.3\%
  & 1.6 & $<0.1\%$ & 6.0 & 77\% & 21 & 18\% \\ \hline Distance
  formula$^{\rm a}$ (\S\ref{sec:dist})& 18 & 43\% & 1.2 & $<0.1\%$ &
  4.5 & 47\% & 14 & 58\% \\ \hline Core radius$^{\rm b}$
  (\S\ref{sec:core})& 21 & 27\% & 0.7 & $<0.1\%$ & 5.4 & 53\% & 19 &
  20\% \\ \hline Magnification bias$^{\rm c}$ (\S\ref{sec:brfn})& 22 &
  23\% & 1.0 & $<0.1\%$ & 4.2 & 40\% & 13 & 62\% \\ \hline Dust
  obscuration$^{\rm d}$ (\S\ref{sec:dust})& 29 & 3.4\% & 1.2 &
  $<0.1\%$ & 5.6 & 77\% & 19 & 33\% \\ \hline
\end{tabular}

\bigskip

{\footnotesize %
  $^{\rm a}$ Self-consistent formula with $\tilde\alpha=0$ is used. \\ 
  $^{\rm b}$ $\xi_{\rm c}=0.2(v/200\,{\rm km\,s^{-1}})^3h^{-1}{\rm
    kpc}$. \\ $^{\rm c}$ Magnification of only the brighter image is
  used. \\ $^{\rm d}$ $\tau_{\rm cV}=1$, $\xi_{\rm d}=0.3h^{-1}{\rm
    kpc}$.  }
\end{table}

As seen from Fig.\ref{fig3} and Table \ref{tab2}, GL statistics based
on theoretical VF's prefers spatially flat low density models with the
non-vanishing cosmological constant (if COBE normalized), as opposed
to the SchVF prediction (this is consistent with the results in
Ref.\cite{T96} which focused on the large separation lensing
statistics using the observed cluster mass function in
Ref.\cite{BC93}). In particular, the model LC matches the observation
fairly well. However, this is originated from the fact that the VF
itself in the model LC underpredict the number of galaxy scale lenses
today (Fig.\ref{fig1}).  Although the KSVF in model LP is the only
theoretical VF (among the four models) consistent with the VF
observation in Fig.\ref{fig1}, Fig.\ref{fig3}b shows that the KSVF
prediction of the model LP is unfavorably larger than the observed
events at small separations. 

As discussed in \S\ref{sec:vfobs} the observed VF on galaxy scales may
be affected by dissipative effects.  However, it is not clear which
the observed VF or the theoretical VF without dissipation is relevant
for GL statistics: if the lensing objects cannot be modeled as
isothermal, then it might well be that the observed velocity
dispersion in their central parts is larger than the effective value
of $v$ inside the Einstein radius of the isothermal model. If this is
the case, the observed VF will overestimate the deflection angle and
so the GL cross section.  In passing we note that, if the observed VF
is affected by dissipative effects, and if theoretical VF's without
dissipation is relevant for GL statistics, then the model LC may
become consistent with both of the VF and GL observations.


\section{Various uncertainties in GL statistics}\label{sec:unc}

Within the SIS lens model and the standard distance formula
($\tilde\alpha=1$), we could not find a cosmological model which
satisfies both GL and VF observations. In this section we consider
various uncertainties (free parameters) in the predictions of GL
statistics. The results are summarized in Table \ref{tab2}

\subsection{Distance formula}\label{sec:dist}

We have no knowledge at present as to what value of the smoothness
parameter $\tilde\alpha$ in the distance formula (Eq.[\ref{40n}])
describes our universe best. First of all, it is important to realize
that the distances appearing in the lens equation (Eq.[\ref{3}]) and
the GL probability formula (Eq.[\ref{12}]) has slightly different
meanings.  First let us discuss why the filled beam ($\tilde\alpha=1$)
distances are used in Eq.(\ref{12}).

Actually Eq.(\ref{12}) has two different derivations. One is the
``self-consistent'' formula\cite{ES} which uses $\tilde\alpha=1$ in
the GL probability formula (as in Eq.[\ref{12}]), and the other is the
``optical-depth'' formula\cite{PG,TOG} which do not distinguish the
distances in Eqs.(\ref{3}) and (\ref{12}), and use the same value of
$\tilde\alpha$ in both the equations. In the self-consistent formula,
the lensing cross section is defined in the source plane (as in
Eq.[\ref{8}]) and the GL probability is defined by the ratio of the
total lensing cross section to the whole area of survey regions in the
source plane. In this case, the distances in Eq.(\ref{12}) measure the
proper area in the lens and source planes subtended by the solid angle
of the regions. Thus, if one observes GL frequency in the whole $4\pi$
steradians in the sky, the filled beam distances should be used in
Eq.(\ref{12}) because (we assume) the universe is on average
homogeneous. On the other hand, in the optical-depth formula, the
lensing cross section is defined in the lens plane and the GL
probability is defined by the mean number of deflections along
line-of-sight. Since $D_1(z_{\rm s}) / D_{\tilde\alpha} (z_{\rm
  s})\leq 1$, the optical depth formula is always smaller than the
self-consistent formula\cite{ES,FFKT} and the difference becomes
larger for smaller $\tilde\alpha$, larger $z_{\rm s}$, $\Omega_0$ and
$\lambda_0$. For example, when $\tilde\alpha=0$,
$(\Omega,\lambda)=(1,0)$, and $z_{\rm s}=2$, the two GL probabilities
differ by a factor of 0.645. We adopt the self-consistent formula
because, as pointed out in Ref.\cite{ES}, the line-of-sight cannot be
treated as a random variable but the source position on the source
plane can be. Although the GL observation data used in \S\ref{sec:hst}
do not cover the whole $4\pi$ steradians, the survey region is wide
enough to assume the statistical homogeneity\cite{W76,KFT}.

On the other hand, the distances appearing in the lens equation
(Eq.[\ref{3}]) measure the proper lengths in the lens and source
planes on scales of the typical image separation. It is shown
(Sasaki\cite{S93}) that the choice of $\tilde\alpha$ in Eq.(\ref{3})
depends on how clumpy one imagines the universe is on that scale. So
$\tilde\alpha$ in Eq.(\ref{3}) is a statistical quantity and may
depend on the directions in the sky. Sasaki\cite{S93} also argues
that, when one calculates the deflection angle
$\hat\alpha=(d/d\xi)\hat\psi$, the density profile of lensing objects
should be subtracted by $\tilde\alpha\bar\rho$ for consistency because
the lens potential $\hat\psi$ (Eq.[\ref{2.5}]) is a perturbative
quantity against the homogeneous background. In the case of strong
lensing, however, the subtraction of $\tilde\alpha\bar\rho$ makes
little difference: if we assume that the lensing object has virialized
before the light passes by it, the density of the lens is at least
$\vartheta_{\rm v}$ times larger than the background when the light is
deflected. Since the deflection angle is roughly proportional to the
mass of the lens, and since $\vartheta_{\rm v}\geq 18\pi^2$
(Eq.[\ref{c18}]), the fractional change in the deflection angle due to
$\tilde\alpha\bar\rho$ is at most a factor of $10^{-2}$, which is
observationally negligible.

To be pedantic, setting $\tilde\alpha=1$ in Eq.(\ref{3}) is
inconsistent because the strong GL cannot occur in the
$\tilde\alpha=1$, exactly homogeneous universe. As $\tilde\alpha$
decreases from unity, the GL probability (without magnification bias)
increases through the geometrical effect: because of the decreasing
focusing, the proper area of the GL cross section increases. The
effect of $\tilde\alpha$ is larger for larger $\Omega_0$ or
$\lambda_0$ universes. This is because $\tilde\alpha$ in the Ricci
focusing term (Eq.[\ref{40n}]) is multiplied by $\Omega(z)$ and larger
$\lambda_0$ makes $\Omega(z)$ approach unity faster at higher
redshifts. On the contrary, if the magnification bias is taken into
account, the GL probability decreases as $\tilde\alpha$ decreases,
because of the factor $\bar\mu:=[D_{\tilde\alpha}(z_{\rm s}) /
D_1(z_{\rm s})]^2$ in Eqs.(\ref{45}) and (\ref{46}). However, this
decrease of GL probability with $\tilde\alpha$ is rather artificial:
intuitively, as $\tilde\alpha$ decreases, the universe becomes clumpy
so the number of lensing objects and the GL probability should
increase. It may be that the number density of lensing objects should
be multiplied by $(1-\tilde\alpha)$ to take this effect into account.

\begin{figure}[htb]
\epsfxsize=8cm\centerline{\epsfbox{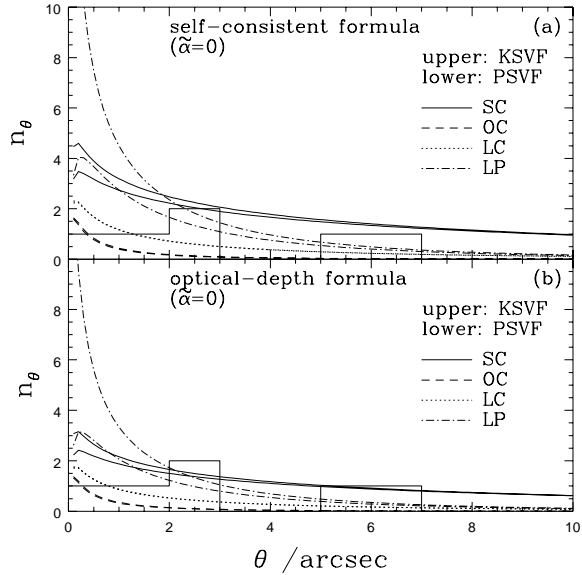}}
\caption{Same as Fig.\protect\ref{fig3}b, except that the
  smoothness parameter (Eq.[\protect\ref{40n}]) is
  $\tilde\alpha=0$. (a) self-consistent formula
  (Eq.[\protect\ref{26}]) with the empty-beam distance
  ($\tilde\alpha=0$) is used; (b): optical-depth formula with the
  empty-beam distance ($\tilde\alpha=0$) is used in stead of
  Eq.(\protect\ref{26}).}
\label{fig4}
\end{figure}

Fig.\ref{fig4} is the same as Fig.\ref{fig3}b except that
$\tilde\alpha=0$ (extreme case). Fig.\ref{fig4}a shows the PSVF and
KSVF predictions from the self-consistent formula, while
Fig.\ref{fig4}b uses the optical depth formula. Because of the
decrease of the magnification bias mentioned above, the GL predictions
decrease as compared with Fig.\ref{fig3}. In our opinion, previous
work (e.g., Ref.\cite{FFKT}) might overestimate the magnification bias
because of the absence of the factor $\bar\mu$ in Eq.(\ref{45}): what
is essential to the magnification bias is the relative magnification
$\mu(y) / \bar\mu$ of a lensed source to unlensed sources. The
$\chi^2$ significance levels of the models are tabulated in the second
row of Table \ref{tab2}. In extremely inhomogeneous ($\tilde\alpha=0$)
universes the models SC and LP are viable.


\subsection{Core radius}\label{sec:core}

It is known\cite{HK} that a small but finite core radius can change
the lensing probability by an order of magnitude: since the strength
of a lens is determined by its central condensation of the surface
density, the removal of the central singularity reduces the lensing
cross section significantly [$\propto (1-x_{\rm c}^{2/3})^3$;
Eq.(\ref{11})].  However, as noted in Ref.\cite{K96}, the inclusion of
core radius increases the magnification bias: the significant decrease
in the GL probability with the core radius is compensated to some
extent by the increase of the magnification bias. The probability even
increases with core radius when the source luminosity is greater than
$\sim 20L_*$.

From Eq.(\ref{7n}), the scaled core radius $x_{\rm c}$ is written as
\begin{equation}\label{40.1}
  x_{\rm c} \simeq 0.72 \left[\frac{v}{100\,{\rm
        km\,s^{-1}}}\right]^{-2} \frac{D_{\rm OS}}{D_{\rm LS}}
  \left[\frac{hD_{\rm OL}}{{\rm Gpc}}\right]^{-1} \frac{h\xi_{\rm
      c}}{{\rm kpc}}\,.
\end{equation}
On galaxy scales the core radius is rather small. Fukugita \&
Turner\cite{FT} estimate $\xi_{\rm c}=324\,h^{-1}{\rm pc}$ at
$v=225\,{\rm km\,s^{-1}}$ from the observation of E galaxies in
Ref.\cite{L85}, which corresponds to $x_{\rm c}\sim 0.12$ for a lens
redshift of $z\sim 0.5$. From the same data, Krauss \& White\cite{KW}
obtains $\xi_{\rm c}=557\,h^{-1}{\rm pc}$ at $v=306\,{\rm km\,s^{-1}}$
($x_{\rm c}\sim 0.14$) and $\xi_{\rm c}\propto L^{0.73}$, though the
scatter in the data is very large.  Using the Faber-Jackson relation,
one obtains $p\sim 2.9$ (see Eq.[\ref{18}]).  Wallington \&
Narayan\cite{WN} argues that $\xi_{\rm c}$ must be smaller than about
$200\,h^{-1}{\rm pc}$ at $v=225\,{\rm km\,s^{-1}}$ ($x_{\rm c}\sim
0.08$) so that the observed even number of lensed images is explained
by the demagnification of the central image. On cluster scales the
core radius is typically $\xi_{\rm c}\sim 20$--$30h^{-1}{\rm kpc}$ at
$v=1000\,{\rm km\,s^{-1}}$ which corresponds to $x_{\rm
  c}\sim0.4$--0.6\cite{FLO,T96,NB}. These data suggest that $p\sim3$
and $\xi_{\rm c}\sim 0.2h^{-1}{\rm kpc}$ at $v=200\,{\rm km\,s^{-1}}$.

\begin{figure}[htb]
\epsfxsize=12cm\centerline{\epsfbox{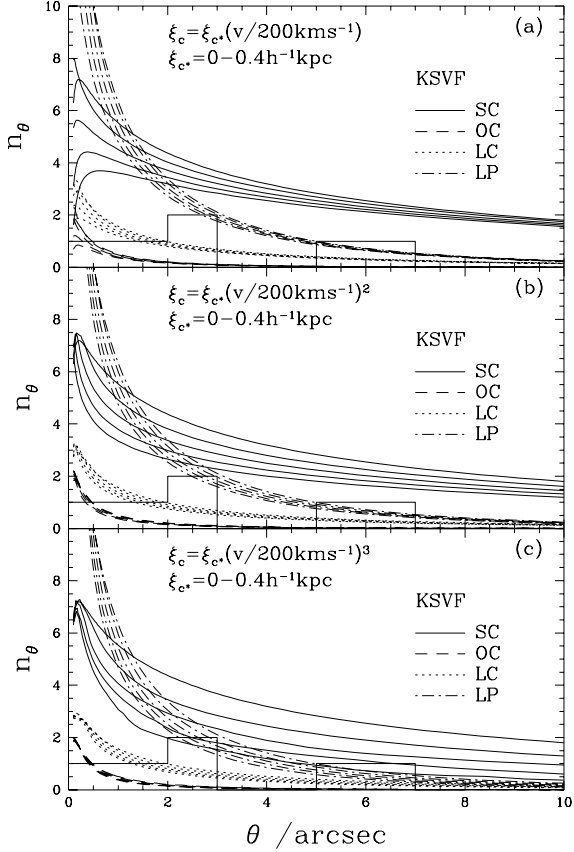}}
  \caption{Same as Fig.\protect\ref{fig3}b, except that the finite
    core radius $\xi_{\rm c}$ is included and only the KSVF
    predictions are shown. (a): $\xi_{\rm c}=\xi_{\rm c*}(v/200\,{\rm
      km\,s^{-1}})$; (b): $\xi_{\rm c}=\xi_{\rm c*} (v/200\,{\rm
      km\,s^{-1}})^2$; (c): $\xi_{\rm c}=\xi_{\rm c*}(v/200\,{\rm
      km\,s^{-1}})^3$; with $\xi_{\rm c*}=0, 0.1, 0.2, 0.3, 0.4
    h^{-1}{\rm kpc}$ from upper to lower curves.}
  \label{fig5}
\end{figure}

Fig.\ref{fig5} plots the same quantities as in Fig.\ref{fig3}b except
that the finite core radius is included and only the KSVF predictions
are shown. We examined $p=1,2$ and 3 (Eq.[\ref{18}]), in
Figs.\ref{fig5}a, b, and c, respectively. From upper to lower of the
five curves for each model, the core radius is $\xi_{\rm c}=0, 0.1,
0.2, 0.3, 0.4 h^{-1}{\rm kpc}$ at $v=200\,{\rm km\,s^{-1}}$. When
$p<2$, the GL probability decreases mainly at small $\theta$ and
asymptotes to the SIS ($\xi_{\rm c}=0$) case at large $\theta$, simply
because $x_{\rm c}\propto v^{p-2}$ (Eq.[\ref{10}]). When $p>2$ the
converse behaviour occurs, and when $p=2$, it decreases uniformly. We
note that the effect of the core radius is slightly smaller in larger
$\lambda_0$ universes. This is because the scale length $\xi_*$
(Eq.[\ref{7n}]) increases with increasing $\lambda_0$ due to the
cosmological geometry: larger $\xi_*$ means that the light traverses
farther away from the lensing object, so the core radius looks
effectively smaller. For the core radius of $\xi_{\rm
  c}=0.2(v/200\,{\rm km\,s^{-1}})^3h^{-1}{\rm kpc}$, the $\chi^2$
significance levels of the models are summarized in the third row of
Table \ref{tab2}.

The above estimates for the core radius of early-type galaxies are
based on pre-HST data. But recent observation of these galaxies by
HST\cite{BYU,GEB} shows that they have nearly singular
cores. Moreover, Ref.\cite{K94} argues that the inclusion of finite
core radius should increase the velocity dispersion of dark matter for
consistency. Based on these arguments, the effect of the core radius
may be smaller than that given in Table \ref{tab2} in practice.

\subsection{Magnification bias}\label{sec:brfn}

The magnification bias (Eqs.[\ref{45}] and [\ref{46}]) contains
$\mu(y)$ which is defined in \S\ref{sec:lensmod} as the total
magnification of all the images. However, depending on the properties
of a QSO sample or the GL configuration, $\mu(y)$ in the bias factor
should be interpreted as the magnification of only the brighter image
or the fainter image among the outer two images. If individual QSO's
in a sample are not examined so closely whether they are lensed or
not, the magnification of only the fainter image should be
used\cite{ST} because the fainter image should be bright enough to be
recognized as one of the multiple images. On the contrary, if one
examines the QSO's closely enough to search for the second image by a
follow-up observation (as in the HST snapshot survey), the
magnification of the brighter image or the total magnification should
be used\cite{CGOT}. Which magnification -- brighter image or the total
-- should be used may depend on the image separation angle: if one
searches for lensed sources of small separation angles, then the total
magnification may be relevant, because it is likely that the
brightness of a lensed source with a small separation is recognized as
the total brightness of all the images.

\begin{figure}[htb]
\epsfxsize=8cm\centerline{\epsfbox{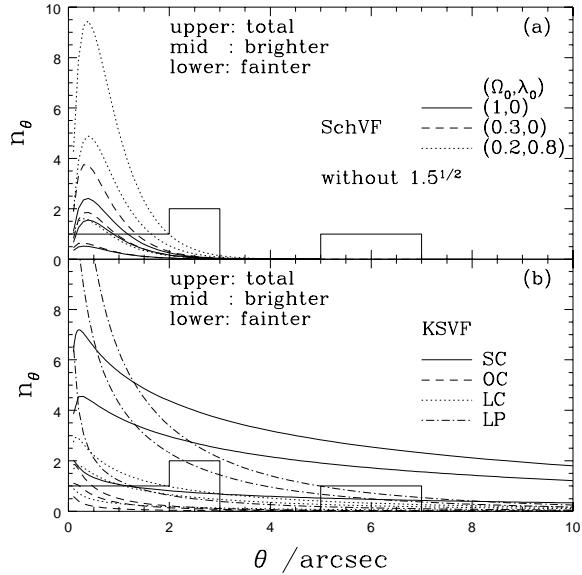}}
  \caption{Same as Fig.\protect\ref{fig3}, except that
    $\mu(y)$ in the magnification bias (\protect\ref{45}) is defined
    as the magnification of the brighter image (the middle curves) or
    the fainter image (lower curves) of the outer two images. The
    upper curves use the total magnification of all the images for
    $\mu(y)$ as in Fig\protect\ref{fig3}.}
  \label{fig6}
\end{figure}

Fig.\ref{fig6} shows how different choices for the magnification bias
affect the GL prediction for the HST survey.  The upper curves for
each model use the total magnification for $\mu(y)$, while the middle
and lower curves use the magnification of brighter and fainter image,
respectively. Only the KSVF predictions and SchVF predictions without
the $1.5^{1/2}$ factor are shown. For the middle curves, the $\chi^2$
significance levels of the models are shown in the fourth row of Table
\ref{tab2}. Because the effect of the magnification bias is very large
in the GL prediction, the different choices of the bias factor changes
the result significantly.

The uncertainties in the QSO luminosity function (Eq.[\ref{47}]),
$\alpha=3.79\pm 0.15$, $k_{\rm L}=3.15\pm 0.10$ and $M^*_0=-20.91\pm
0.25+5\log h$, change the GL prediction by $\sim\pm 40\%$, $\pm 40\%$
and $\pm 20\%$, respectively, while $\beta=1.44\pm 0.20$ does not
cause significant changes because QSO's brighter than $L_*$ are
preferentially selected in the HST snapshot survey in order to achieve
high efficiency of the lens detection\cite{K91}.


\subsection{Dust obscuration}\label{sec:dust}

When the light from a lensed source passes by the lensing galaxy, the
lensed image may be obscured by the dust ingredients in the
galaxy. For example, the scale length (Eq.[\ref{7n}])
\begin{equation}
  \xi_* \simeq 1.4h^{-1}{\rm kpc} \left[\frac{v}{100\,{\rm
        km\,s^{-1}}}\right]^2 \frac{D_{\rm LS}}{D_{\rm OS}}
  \frac{hD_{\rm OL}}{\rm{Gpc}}
\end{equation}
is about $2h^{-1}{\rm kpc}$ for a lensing galaxy of $v=200\,{\rm
  km\,s^{-1}}$ and $z=0.5$, which may be well inside the baryonic
parts of the galaxy. The dust obscuration reduces the GL probability
in two ways: i) due to the dust demagnification, the magnification
bias decreases; ii) brightness ratio of the images is amplified and so
it becomes hard to detect the second image, because the fainter image,
which is nearer to the lens center, is more appreciably demagnified
than the brighter image.

Let us estimate this effect quantitatively by a simple
model\cite{FP,T93,K96}. Since elliptical galaxies dominate over spiral
galaxies in GL statistics (\S\ref{sec:schvf}), the former is
considered here as lensing galaxies. Let $\tau(\xi)$ be the
(2-dimensional) optical depth profile due to obscuration by dusts at
the impact parameter $\xi$ from the center of the galaxy. We assume
the following form of $\tau(\xi)$:
\begin{equation}\label{49}
  \tau(\xi)=\tau_{\rm c}/[1+(\xi/\xi_{\rm d})^2]
\end{equation}
where $\tau_{\rm c}$ is the optical depth at the center and $\xi_{\rm
  d}$ is the dust core radius. The light ray which passes by the
lensing galaxy at $\xi$ is demagnified by a factor $e^{ -\tau(\xi)}$.
Accordingly we have to redefine the total magnification $\mu(y)$ and
the brightness ratio $r(y)$ in Eq.(\ref{45}) by replacing $\mu_{\rm
  p}(x)$ in \S\ref{sec:lensmod} with $\mu_{\rm p}(x) e^{ -\tau(\xi_*
  x)}$, and recalculate the magnification bias. The optical depth is
sensitive to the wavelength $\lambda$. Following Ref.\cite{T93}, we
use the fitting formula of Ref.\cite{S79}:
\begin{equation}
  \tau_{\rm c}(\lambda) \simeq \tau_{\rm cV}X(x)/3.2\,,
\end{equation}
where $x:=(\lambda/\mu{\rm m})^{-1}$, $\tau_{\rm cV}$ is the optical
depth in V-band ($x_{\rm V}=1.8$), and the functional form of $X(x)$
is found in Ref.\cite{T93}. If QSO is observed in the wavelength
$\lambda_{\rm obs}$ and the lensing galaxy is at redshift $z$, then
the demagnification factor should be evaluated at the wavelength
$\lambda_{\rm obs}/(1+z)$. Wise \& Silva\cite{WS96} obtain $\tau_{\rm
  cV}\sim 1$ comparing their dust model with the data of 52 elliptical
galaxies. We estimate $\xi_{\rm d}\sim 0.3h^{-1}{\rm kpc}$, assuming
that $\xi_{\rm d}$ is 0.1 times the effective radius of the de
Vaucouleurs profile\cite{BT}. To be more realistic, $\tau_{\rm cV}$
and $\xi_{\rm d}$ should be dependent on the velocity dispersion $v$
(the sizes of galaxies) and on $z$ (the evolution of
galaxies). However, we set them constant for simplicity because we aim
at examining the extent to which the dust obscuration affects the GL
probability.

\begin{figure}[htb]
\epsfxsize=8cm\centerline{\epsfbox{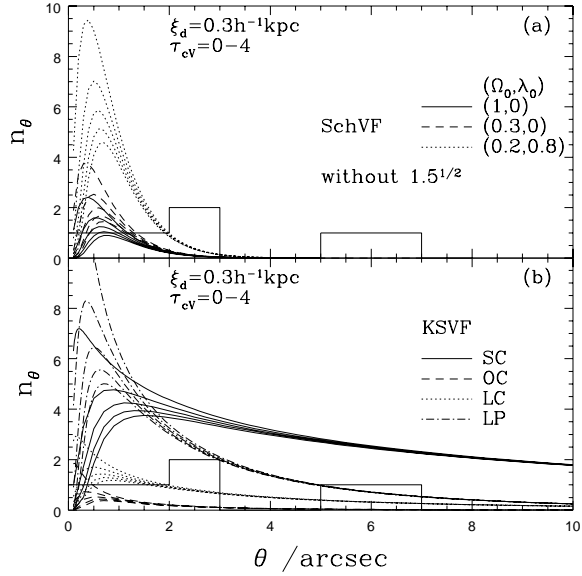}}
  \caption{Same as Fig.\protect\ref{fig3}, except that the dust
    obscuration effect by lensing galaxies is included, and only KSVF
    predictions and SchVF predictions without the $1.5^{1/2}$ factor
    are shown. From upper to lower curves, the optical depth of the
    dust obscuration is $\tau_{\rm cV}=0,1,2,3,4$ (see
    Eq.[\protect\ref{49}])}
  \label{fig7}
\end{figure}

Fig.\ref{fig7} shows the dust obscured GL predictions on the HST
survey. Only KSVF predictions and SchVF predictions without the
$1.5^{1/2}$ factor are shown. Among the five curves for each model,
the optical depth is $\tau_{\rm cV}=0,1,2,3,4$ from the upper to lower
curves, and $\xi_{\rm d}$ is set to $0.3h^{-1}{\rm kpc}$. The dust
obscuration effect is considerably large at the small separation
angles. We note, however, that the effect becomes small if one
includes a finite core radius, because the brightness ratio becomes
small. For the ``realistic'' value $\tau_{\rm cV}=1$, the $\chi^2$
significance levels of the models are presented in the fifth row of
Table \ref{tab2}.

Let us note general features of the effect of the dust
obscuration. Larger $\lambda_0$ universes have smaller effect of the
dust obscuration because the light rays are likely to pass by farther
from the lens center, as noted in \S\ref{sec:core}. If $\xi_{\rm d}$
increases with $v$ less rapidly than $\propto v^2$, the GL probability
decreases mainly at small separation and asymptotes to the no-dust
curve at large separation, and the converse is true if $\xi_{\rm d}$
increases more rapidly than $v^2$, for exactly the same reason as in
Fig.\ref{fig5}. If $\tau_{\rm cV}$ increases with $z$ due to the star
formation activities at higher redshifts\cite{FP}, lensed images with
small separations are more obscured than those with large separations,
because the former are likely to be produced by lensing objects of
high redshifts (see Eq.[\ref{10}]). Observation\cite{WS96,G96} suggest
that the dusts are more diffusely distributed as $\rho_{\rm
  dust}(r)\propto r^{-1}$ rather than as Eq.(\ref{49}). In this case
large separation images are more obscured than in Fig.\ref{fig7}.


\section{Predictions for future surveys}\label{sec:dss}

With the limited statistics of the current data, the constraints on
the parameters are not yet decisive. We expect, however, that a much
larger number of sources will be homogeneously sampled in one
systematic lens survey in the near future. For example, the Sloan
Digital Sky Survey (SDSS) plans a spectroscopic survey of $10^5$ QSO's
over $\pi$ steradians brighter than 19 magnitude in
g-band\cite{GW}. In this section, we will make predictions on future
GL surveys, bearing mainly the SDSS in mind.

Let $n_{\theta}(\theta,S)d\theta$ be the expected number of lensed
QSO's with image separation $\theta \sim \theta+d\theta$ and observed
flux brighter than $S$, within the solid angle $\Omega_{\rm sa}$ in
the sky. We assume that the survey examines all the QSO's in the sky
homogeneously. Then, using the cumulative QSO luminosity function
${\mit\Phi}_{\rm Q}$ in \S\ref{sec:magb}, $n_{\theta}(\theta,S)$ can
be calculated as
\begin{equation}\label{44n}
  n_{\theta}(\theta,S) = \Omega_{\rm sa} \int_0^{z_{\rm max}} dz
  \frac{dr}{dz} r^2(z) {\mit\Phi}_{\rm Q}(L,z) P^{\rm
    B}_{\theta}(\theta,S,r_*,z)\,,
\end{equation}
where $L$ is related to $S$ by Eq.(\ref{45.5}), $r(z):=(1+z)D_1(z)$ is
the proper motion distance from us to $z$, and Eqs.(\ref{20}) and
(\ref{46}) are substituted for $P^{\rm B}_{\theta}$. Since we use
Eq.(\ref{47}) for ${\mit\Phi}_{\rm Q}(L)$ which assumes
$(\Omega,\lambda)=(1,0)$ and $\gamma=0.5$, these values of the
parameters are also used in Eq.(\ref{45.5}) and $r(z)$. We set $z_{\rm
  max}=5$\cite{WN}, and use the same detection limit
$r_*=19(\theta/{\rm arcsec})^{0.85}$ as that of the HST survey.

\begin{figure}[htb]
\epsfxsize=8cm\centerline{\epsfbox{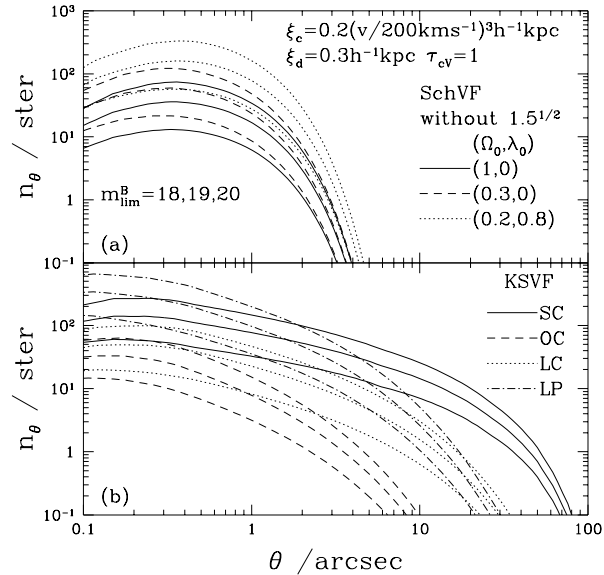}}
  \caption{Image separation $\theta$ distribution of the expected
    number of all the lensed QSO's per unit steradian in the sky
    (Eq.[\protect\ref{44n}]). (a): predictions from SchVF. (b):
    predictions from KSVF. From lower to upper of the three curves for
    each model, the limiting magnitude is $m=18,19,20$ in B-band.}
  \label{fig8}
\end{figure}

\begin{table}[htb]
\caption{The expected number of lensed QSO's per unit steradian
  brighter than 18, 19 and 20 B-mag. (a): SchVF predictions within the
  separation range $0'' < \theta < 5''$; (b): PSVF and KSVF
  predictions within $0'' < \theta < 5''$; (c): PSVF and KSVF
  predictions within $5'' < \theta < 10''$; for models in Table
  \protect\ref{tab1}. We adopt $\xi_{\rm c}=0.2(v/200\,{\rm km\,s^{-1}})^3h^{-1}{\rm
    kpc}$, $\tau_{\rm cV}=1$ and $\xi_{\rm d}=0.3h^{-1}{\rm kpc}$ as
  the models of the core radius and the dust obscuration (see
  \S\protect\ref{sec:core} and \S\protect\ref{sec:dust}).}
\label{tab3}

\bigskip

\begin{tabular}{l|ccc|ccc}
\multicolumn{7}{l}{$0''<\theta<5''$} \\ \hline
(a) & \multicolumn{3}{c|}{SchVF without $1.5^{1/2}$} &
\multicolumn{3}{c}{SchVF with $1.5^{1/2}$} \\
$(\Omega_0, \lambda_0)$ & $B<18$ & 19 & 20 & 18 & 19 & 20 
\\ \hline
(1,0)& 13 & 35 & 70 & 25 & 66 & 132 \\
(0.3,0)& 19 & 52& 106 & 37 & 100 & 202 \\
(0.2,0.8)& 57 & 158 & 326 & 111 & 306 & 629 \\ \hline
\end{tabular}

\medskip

\begin{tabular}{l|ccc|ccc}
\multicolumn{7}{l}{$0''<\theta<5''$} \\ \hline
(b) & \multicolumn{3}{c|}{PSVF} & \multicolumn{3}{c}{KSVF} \\
Model & $B<18$ & 19 & 20 & 18 & 19 & 20 \\ \hline
SC& 102 & 233 & 436 & 118 & 270 & 508 \\
OC& 11 & 27 & 53 & 10 & 25 & 48 \\
LC& 28 & 74 & 151 & 27 & 72 & 146 \\
LP& 65 & 172 & 348 &114 & 305 & 620 \\ \hline
\end{tabular}

\medskip

\begin{tabular}{l|ccc|ccc}
\multicolumn{7}{l}{$5''<\theta<10''$} \\ \hline
(c) & \multicolumn{3}{c|}{PSVF} & \multicolumn{3}{c}{KSVF} \\
Model & $B<18$ & 19 & 20 & 18 & 19 & 20 \\ \hline
SC& 43 & 92 & 165 & 46 & 97 & 176 \\
OC& 0.5 & 1.0 & 1.9 & 0.4 & 0.9 & 1.6 \\
LC& 5.6 & 14 & 28 & 5.4 & 14 & 27 \\
LP& 10 & 25 & 49 & 12 & 30 & 58 \\ \hline
\end{tabular}

\end{table}

Fig.\ref{fig8} and Table \ref{tab3} show the predictions on future
surveys: the expected number of lensed QSO's per unit steradian
($\Omega_{\rm sa}=1$). We adopt $\xi_{\rm c}=0.2(v/200\,{\rm
  km\,s^{-1}})^3h^{-1}{\rm kpc}$, $\tau_{\rm cV}=1$ and $\xi_{\rm
  d}=0.3h^{-1}{\rm kpc}$ as the models of the core radius and the dust
obscuration, which we regard as the realistic values (see
\S\ref{sec:core} and \ref{sec:dust}). The standard angular diameter
distance ($\tilde\alpha=1$) is used. In Fig.\ref{fig8}, the KSVF
predictions and the SchVF predictions without the $1.5^{1/2}$ factor
are shown. Among the three curves for each model, the apparent
limiting magnitudes corresponding to $S$ are $m=18,19,20$ in B-band
from lower to upper curves. Table \ref{tab3}a shows the SchVF
predictions within the separation range $0''< \theta <5''$, while
Table \ref{tab3}b and \ref{tab3}c show the PSVF and KSVF predictions
within $0''<\theta<5''$ and $5''< \theta <10''$. The KSVF prediction
is strongly dependent on $\Omega_0$ and the bias parameter
$b$. Therefore, we believe that Fig.\ref{fig8}b will provide a
promising test for one of these parameters if the other is given
independently.

We propose that the cosmological tests should be carried out at a
large separation range so that various uncertainties do not enter: the
dust obscuration may be ineffective at $\theta\simgt 3''$
(Fig.\ref{fig7}); theoretical VF is free from uncertainties concerning
the formation rate or the dissipative processes (\S\ref{sec:vfobs}) at
$\theta\simgt 7''$ (corresponding to $v\sim 500\,{\rm km\,s^{-1}}$;
Eqs.[\ref{10}] and [\ref{40.1}]) which nearly discriminate the galaxy
scale and cluster scale; etc. Alternatively, the lensing events with
$\theta\simlt 7''$ should provide information about VF and the
dissipative effects on the dynamics of galaxy formation.

In closing we give some cautions on Fig.\ref{fig8}. It may be that the
predictions in the figure are overestimated because of the assumption
that the QSO luminosity function of Boyle et al.\cite{BSP} is valid up
to $z=3$. The dependences on the core radius and the dust obscuration
are similar to Figs.\ref{fig5} and \ref{fig7}. The uncertainties in
the values of $M^*_0$ and $k_L$ change the predictions by $\sim \pm
20\%$ and $\pm 10\%$. Those of $\alpha$ and $\beta$ do not cause a
significant change.


\section{Summary and conclusion}\label{sec:sum}

To date, it is well known that GL statistics disfavors cosmological
models dominated by the positive cosmological constant $\Lambda$
\cite{FT,MR,K96}, in contrast to the other cosmological tests
suggesting low density universe with non-zero
$\Lambda$\cite{SU93,LLVW}. In the light of this, we reexamined the
current constraints on $\Lambda$ from GL statistics, mainly paying
attention to the role of VF in the calculation of GL probability.

Most previous work on GL statistics uses SchVF which counts only
luminous galaxies as lensing objects. Moreover constant comoving
number of lenses (no evolution) is often assumed. To take account of
dark lensing objects and their number evolution, VF is constructed
theoretically from the PS theory in
\S\ref{sec:vf}\cite{NW,K95}. Actually the PS theory in its original
form is not a perfect one in deriving VF, and we attempted in
\S\ref{sec:ksvf} to construct a more realistic VF (KSVF) from the
hierarchical clustering model in Ref.\cite{KS}. Then we compared these
theoretical VF's with the observation of VF by Ref.\cite{SH93} for
COBE normalized CDM and PIB models in \S\ref{sec:vfobs}. There we
found that KSVF in the model LP -- the spatially flat PIB model with
($\Omega_0$, $\lambda_0)=(0.2, 0.8$) (Table \ref{tab1}) -- matches the
observation best (Fig.\ref{fig1}). We also discussed that the model LC
-- the CDM model with $(0.2, 0.8$) -- may become viable if the
observed VF on galaxy scales is affected by the dissipative processes.

At first sight, one might expect that the inclusion of dark lenses and
their evolution into GL statistics would lead to tighter limits on
$\lambda_0$. However, when we compared the predictions of the
theoretical VF's with the HST snapshot survey\cite{M93b} in
\S\ref{sec:hst}, we saw that spatially flat models with non-zero
$\Lambda$ (model LC) are viable among the COBE normalized models
(Fig.\ref{fig3}). This is because the PSVF in low density universes
underpredicts the number of galaxy scale lenses
(Fig.\ref{fig1}). Though the KSVF in the model LP is consistent with
the VF observation, the model predicts higher lensing frequency at
small separations than observed. We could find no COBE normalized
cosmological model which satisfies both VF and GL observations, within
the SIS, $\tilde\alpha=1$ and no dust model.

In \S\ref{sec:unc}, we examined various uncertainties in the GL
predictions systematically: the choice of the $\tilde\alpha$ parameter
in the distance formula (Eq.[\ref{40n}]), core radius $\xi_{\rm c}$,
magnification bias, and the obscuration of images by dusts in the
lensing galaxies. All these uncertainties reduce the GL prediction
considerably from the standard calculation within a range of the
realistic values of free parameters. The core radius and the dust
obscuration are effective mainly at large and small separation range,
respectively (Figs.\ref{fig5} and \ref{fig7}), while the uncertainties
in $\tilde\alpha$ and the magnification bias reduce the GL prediction
uniformly on all separation range. Because of these uncertainties, it
is premature to put strong constraints on cosmological models,
including $\Lambda$, from the rather small number of observed lensed
sources. When a large number of QSO's are homogeneously sampled, a
more reliable comparison will become possible. See Table \ref{tab2}
for a summary.

In \S\ref{sec:dss}, we made GL predictions relevant for the next
generation survey like the SDSS. We believe that the predictions from
theoretical VF will provide a promising test for $\Omega_0$ and
$\lambda_0$, combining with the COBE normalization. We proposed that
the comparison between the predictions and the data should be made at
large separation so that the various uncertainties do not enter:
$\theta\simgt 7''$ may be ideal for cosmological tests because the
theoretical VF is expected to be reliable on large scales. Lensing at
the intermediate range ($3''\simlt \theta \simlt7''$) may provide
information about the degree of dissipative effects on the galaxy
formation. At the same time, the theory of VF needs improvements on
the formation rate (\S\ref{sec:ksvf}) as well as on the dissipative
effects.

\section*{Acknowledgements}

We would like to thank K.~Shimasaku and D.~Maoz for providing the
observational data, N.~Sugiyama for permitting us to use his numerical
program for the COBE normalization, and E.D.~Stewart for careful
reading of the manuscript. TTN would like to thank S.~Sasaki,
T.~Kitayama and K.~Ikumi for helpful discussions, and gratefully
acknowledges the support from JSPS fellowship.  This work is supported
in part by grants-in-aid by the Ministry of Education, Science, Sports
and Culture of Japan (4125, 07CE2002).


\appendix

\section{Specific solutions of the Dyer-Roeder
  equation}\label{app:dist}

In this appendix, the specific solutions of the Dyer-Roeder equation
(Eq.[\ref{40n}]) are presented in the cases $\tilde\alpha=1$ (filled
beam) and $\tilde\alpha=0$ (empty beam). We use the units $c=H_0=1$,
and the notations $a:=(1+z)^{-1}$ and
\begin{equation}\label{a1}
w:=\Omega^{-1}-1 = \left\{
\begin{array}{ll}
  w_0 a \quad & (\lambda=0) \\ w_0 a^3 \quad & (\Omega+\lambda=1)
\end{array}\right.
\end{equation}
with $w_0:=\Omega_0^{-1}-1$ (see also Eq.[\ref{1}]).\\ i)
$\tilde\alpha=1$ (filled beam)\\ i-i) $\lambda=0$
\begin{equation}
  D_1(z_1,z_2)=\frac{2 a_2}{\sqrt{\Omega_0}}
  \left[a_1^{1/2}(1+w_1)^{1/2}(1+2w_2) - a_2^{1/2}(1+w_2)^{1/2}
    (1+2w_1)\right]
\end{equation}
i-ii) $\Omega+\lambda=1$
\begin{equation}\label{a3}
  D_1(z_1,z_2) = \frac{2 a_2}{\sqrt{\Omega_0}}
  \left[a_1^{1/2}F\left(\frac12,\frac16,\frac76;-w_1\right) -
    a_2^{1/2}F\left(\frac12,\frac16,\frac76;-w_2\right) \right]\,,
\end{equation}
where $F$ is the hypergeometric function of type (2,1).\\
ii) $\tilde\alpha=0$ (empty beam)\\ ii-i) $\lambda=0$
\begin{eqnarray}
  D_0(z_1,z_2) = && \frac1{8\sqrt{\Omega_0}w_0^2a_1} \Bigl[
  2a_1^{1/2}(1+w_1)^{1/2}(2w_1-3) \nonumber \\ && -
  2a_2^{1/2}(1+w_2)^{1/2}(2w_2-3) + 3|w_0|^{-1/2}(\eta_1 - \eta_2)
  \Bigr]\,,
\end{eqnarray}
where $\eta:={\rm arccosh}(1+2w)$ when $\Omega<1$ and
$\eta:=\arccos(1+2w)$ when $\Omega>1$.\\ 
ii-ii) $\Omega+\lambda=1$
\begin{equation}\label{a5}
  D_0(z_1,z_2) = \frac2{5\sqrt{\Omega_0}a_1}
  \left[a_1^{5/2}F\left(\frac12,\frac56,\frac{11}6;-w_1\right) -
    a_2^{5/2}F\left(\frac12,\frac56,\frac{11}6;-w_2\right) \right]\,.
\end{equation}
We do not use Eqs.(\ref{a3}) and (\ref{a5}) because the evaluation of
the hypergeometric function is much more time-consuming than the
direct numerical integration of Eq.(\ref{40n}).


\section{Analytic expressions for $v_1$ and $v_2$}\label{app:v1v2}

From Eqs.(\ref{10}) and (\ref{18}), $v_1$ and $v_2$ in Eqs.(\ref{16})
and (\ref{26}) are determined by the equation:
\begin{equation}\label{b1}
  v^4 - X^2v^{2p} - Y^2 = 0\,,
\end{equation}
where
\begin{equation}
  X := \frac1{4\pi} \frac{D_{\rm OS}\xi_{\rm c*}}{D_{\rm OL}D_{\rm
      LS}}\,, \quad Y := \frac{\theta}{8\pi} \frac{D_{\rm OS}}{D_{\rm
      LS}}\,.
\end{equation}
The analytic expressions for $v_1$ and $v_2$ in the cases $p=0,1,2,3$
and 4 are:
\begin{equation}
p=0:\ v_1 = (X^2+Y^2)^{1/4}
\end{equation}
\begin{equation}
  p=1:\ v_1 = [ X^2 + (X^4 + 4Y^2)^{1/2} ]^{1/2}/\sqrt2
\end{equation}
\begin{equation}
  p=2:\ v_1 = \sqrt{Y} (1-X^2)^{-1/4}
\end{equation}
\begin{equation}
  p=3:\begin{array}{l} \displaystyle{ v_1 = \frac1X \left[ \frac13 +
        \frac23\cos\left(\frac23\pi - Z\right) \right]^{1/2},} \\ \\ 
    \displaystyle{ v_2 = \frac1X \left[ \frac13 + \frac23\cos Z
      \right]^{1/2}}
\end{array}
\end{equation}
\begin{equation}
  p=4:\ v_{1,2} = (\sqrt2 X)^{-1/2}[ 1 \mp (1-4X^2Y^2)^{1/2}]^{1/4}\,,
\end{equation}
where
\begin{equation}
  Z := \frac13 \arccos \left[ 1 - \frac{27}2 (X^2Y)^2 \right]\,.
\end{equation}
When $p\leq 2$, $v_2=\infty$.


\section{Spherical collapse in $\Omega+\lambda=1$
  universe}\label{app:sph}

Consider a local spherical region of radius $r$ and mass $M$ in an
$\Omega+\lambda=1$ universe with $\lambda>0$. From the expansion
equation:
\begin{equation}
  \frac{d^2r}{dt^2} = -\frac{M}{r^2} + \frac{\Lambda}3 r\,,
\end{equation}
one obtains
\begin{equation}\label{c2}
  t = \int_0^r dr'\left( 2E + \frac{2M}{r'} + \frac{\Lambda}3r'^2
  \right)^{-1/2}
\end{equation}
when $dr/dt>0$ (we neglect the decaying mode). The sphere ``turns
around'' at a radius $r_{\rm ta}$ defined by
\begin{equation}
  E = -\frac{M}{r_{\rm ta}} - \frac{\Lambda}6 r_{\rm ta}^2
\end{equation}
if
\begin{equation}
  \zeta := \frac{\Lambda r_{\rm ta}^3}{6M} < \frac12
\end{equation}
[our $\zeta$ corresponds to $\eta/2$ of Ref.\cite{LLPR}]. Then
Eq.(\ref{c2}) is rewritten as\cite{H93}
\begin{equation}\label{c5}
  Ht = \left(\frac{\zeta}{\lambda}\right)^{1/2} \int_0^y dx
  \left[\frac1x - (1+\zeta) + \zeta x^2 \right]^{-1/2}\,,
\end{equation}
where $y:=r/r_{\rm ta}$. On the other hand, from Eq.(\ref{2}) the age
of the global universe in $\Omega+\lambda=1$ models is
\begin{equation}\label{c10}
  Ht = \frac13 (\Omega w)^{-1/2}{\rm arccosh}(1 + 2w)\,,
\end{equation}
where $w$ is defined in Eq.(\ref{a1}). The sphere collapses to $r=0$
in twice the turn-around time. In reality, however, it will reach a
virialized state through the violent relaxation\cite{L67} and
thereafter $r$ stays constant $r_{\rm v}$ determined by\cite{LLPR}
\begin{equation}\label{c6}
4\zeta y_{\rm v}^3 -2(1+\zeta) y_{\rm v} + 1 = 0\,,
\end{equation}
where and hereafter {\em the subscript} v {\em indicates the
  virialization time}. Eq.(\ref{c6}) has the solution
\begin{equation}\label{c7}
  y_{\rm v} = \left( \frac{2+2\zeta}{3\zeta} \right)^{1/2} \!\!\! \cos
  \left\{\frac23\pi - \frac13\arccos\left[ -
      \frac1{\zeta}\left(\frac{3\zeta}{2+2\zeta}\right)^{3/2} \right]
  \right\}
\end{equation}
in the range $0<\zeta<0.5$. Let $\rho:=3M/(4\pi r^3)$ be the mean
density inside the sphere. The overdensity against the background is
written as
\begin{equation}\label{c8n}
  \vartheta:=1+\delta := \frac{\rho}{\bar\rho} = \frac{w}{y^3\zeta}\,.
\end{equation}
Let us calculate the overdensity of an virialized object at the
virialization time:
\begin{equation}\label{c8}
  \vartheta_{\rm v} := \frac{\rho_{\rm v}}{\bar\rho_{\rm v}} =
  \frac{w_{\rm v}}{y_{\rm v}^3\zeta}.
\end{equation}
In order to relate $\zeta$ with $w_{\rm v}$, we equate the local time
inside the sphere (Eq.[\ref{c5}]) with the global cosmic age
(Eq.[\ref{c10}]) at the virialization epoch.  The former is often
approximated by the collapse time:
\begin{eqnarray}
  H_{\rm v}t_{\rm v} &=& 2\left(\frac{\zeta}{\lambda_{\rm
        v}}\right)^{1/2} \int_0^1 dx \left[ \frac1x - (1+\zeta) +
    \zeta x^2 \right]^{-1/2} \\ &=& \frac13\lambda_{\rm v}^{-1/2} [
  AK(k) - B{\mit\Pi}(\nu,k) ] \,,
\label{c12}
\end{eqnarray}
where $K(k)$ and ${\mit\Pi}(\nu,k)$ are the complete elliptic
integrals of the first and third kind, respectively (see
Ref.\cite{PTVF} for their definitions), and
\begin{equation}
  A := [1 + (1-2\zeta)^{1/2}]C, \quad B := 2(1-2\zeta)^{1/2}C,
\end{equation}
\begin{equation}
  C := 12\zeta^{-1/2}[2-\zeta+2(1-2\zeta)^{1/2}]^{-1/2}
\end{equation}
\begin{equation}
  k^2 :=
  \frac{2-\zeta-2(1-2\zeta)^{1/2}}{2-\zeta+2(1-2\zeta)^{1/2}}\,,
\end{equation}
\begin{equation}
  \nu := -\zeta^{-2}[1-\zeta-(1-2\zeta)^{1/2}]^2\,.
\end{equation}
Note that Eq.(\ref{c12}) does not assume the spatial flatness
$\Omega+\lambda=1$. From Eqs.(\ref{c10}) and (\ref{c12}), $\zeta$ and
$w_{\rm v}$ are related as
\begin{equation}\label{c17}
  w_{\rm v} = \frac12\{ \cosh[AK(k)-B{\mit\Pi}(\nu,k)] - 1 \}\,.
\end{equation}
Now we obtain the analytic formula for $\vartheta_{\rm v}$ in
$\Omega+\lambda=1$ universe through the parameter $\zeta$. Given a
value of $\zeta$, one can compute $\vartheta_{\rm v}$ and $\Omega_{\rm
  v}$ from Eqs.(\ref{c7}), (\ref{c8}), (\ref{c17}) and (\ref{a1}). In
practice, however, it is useful to write $\vartheta_{\rm v}$ as a
function of $\Omega_{\rm v}$. For this purpose, we give the fitting
formulae:
\begin{equation}\label{c18}
  \vartheta_{\rm v} \simeq 18\pi^2 \Omega_{\rm v}^{-0.573} \quad{\rm
    or}
\end{equation}
\begin{equation}\label{c19}
  \vartheta_{\rm v} \simeq 18\pi^2 (1 + 0.40929 w_{\rm v}^{0.90524} ).
\end{equation}
Eqs.(\ref{c18}) and (\ref{c19}) are accurate within 5\% for
$\Omega_{\rm v}>0.1$ and 1.7\% for $\Omega_{\rm v}>0.01$,
respectively.

Next, let us calculate $\delta_{\rm c}$, the density contrast in the
early universe extrapolated linearly to the virialization time. Since
$y\ll 1$ in the early stages, we expand the integrand of Eq.(\ref{c5})
in powers of $x$ and keep the linear term only:
\begin{eqnarray}
  Ht &=& \left(\frac{\zeta}{\lambda}\right)^{1/2} \int_0^y \sqrt{x} [1
  + \frac12(1+\zeta)x ] dx \\ &=&\frac23
  \left(\frac{\zeta}{\lambda}\right)^{1/2} y^{3/2} [ 1 +
  \frac3{10}(1+\zeta) y ]\,.
\label{c22}
\end{eqnarray}
Similarly, $w\ll 1$ in Eq.(\ref{c10}) yields
\begin{equation}\label{c23}
  Ht = \frac23\Omega^{-1/2}\,.
\end{equation}
Equating Eq.(\ref{c22}) with (\ref{c23}), $y$ is written iteratively
as
\begin{equation}
  y = \left(\frac{w}{\zeta}\right)^{1/3}\left[ 1 -
    \frac15(1+\zeta)\left(\frac{w}{\zeta}\right)^{1/3}\right]\,.
\end{equation}
Substituting this into Eq.(\ref{c8n}), the density contrast of the
spherical region in the early universe is
\begin{equation}\label{c25}
  \delta = \frac35(1+\zeta)\left(\frac{w}{\zeta}\right)^{1/3}
\end{equation}
which grows as $\propto w^{1/3}\propto a$ in accord with the linear
perturbation theory\cite{P80}. The linear growth rate in
$\Omega+\lambda=1$ universe is\cite{M95}
\begin{equation}\label{c26}
  D = w^{1/3} F\left(\frac13,1,\frac{11}6;-w\right)
\end{equation}
which is normalized as $D/a\to w_0^{1/3}$ when $a\to 0$, with $F$
being the hypergeometric function of type (2,1). Extrapolating
Eq.(\ref{c25}) by Eq.(\ref{c26}) until $a_{\rm v}$, one obtains
\begin{equation}\label{c27}
  \delta_{\rm c} = \frac35 F\left(\frac13,1,\frac{11}6;-w_{\rm
      v}\right) (1+\zeta)\left(\frac{w_{\rm v}}{\zeta}\right)^{1/3}\,.
\end{equation}
One can compute $\delta_{\rm c}$ and $\Omega_{\rm v}$ from
Eqs.(\ref{c17}), (\ref{c27}) and (\ref{a1}) through the parameter
$\zeta$. We give fitting formulae so that one can obtain $\delta_{\rm
  c}$ directly from a given value of $\Omega_{\rm v}$:
\begin{equation}\label{c28}
  \delta_{\rm c} \simeq \frac3{20}(12\pi)^{2/3} \Omega_{\rm
    v}^{0.00539} \quad{\rm or}
\end{equation}
\begin{equation}\label{c29}
  \delta_{\rm c} \simeq \frac3{20}(12\pi)^{2/3}(1 +
  0.012299\log_{10}\Omega_{\rm v} ).
\end{equation}
Eqs.(\ref{c28}) and (\ref{c29}) are accurate within 2.4\% for
$\Omega_{\rm v}>0.1$ and 0.1\% for $\Omega_{\rm v}>0.01$,
respectively.

Of more practical use (in, e.g., the PS theory) are the following
quantities
\begin{equation}\label{c29.4}
  \vartheta_{\rm v0} := \rho_{\rm v}/\bar\rho_0 = \vartheta_{\rm v}
  /a_{\rm v}^3
\end{equation}
\begin{eqnarray}\label{c29.5}
  \delta_{\rm c0} &:=& \frac{D_0}{D_{\rm v}}\delta_{\rm c} = \frac35
  F\left(\frac13,1,\frac{11}6;-w_0\right)
  (1+\zeta)\left(\frac{w_0}{\zeta}\right)^{1/3} \\ &=&
  \frac3{20}(12\pi)^{2/3}\frac1{a_{\rm v}}
  F\left(\frac13,1,\frac{11}6;-w_0\right) X(w_{\rm v})\,,
\label{c30}
\end{eqnarray}
where
\begin{equation}\label{c31}
  X(w_{\rm v}) \simeq \Omega_{\rm v}^{-0.215} \quad{\rm or}
\end{equation}
\begin{equation}\label{c32}
  X(w_{\rm v}) \simeq 1+(5.1066 w_{\rm v}^{-1.0812} + 2.0215 w_{\rm
    v}^{-0.35396})^{-1}.
\end{equation}
The accuracies of Eqs.(\ref{c31}) and (\ref{c32}) are within 1.8\% for
$\Omega_{\rm v}>0.1$ and 3.1\% for $\Omega_{\rm v}>0.01$,
respectively. The corresponding formulae in an open ($\lambda=0$)
universe are found in Ref.\cite{LC93}.


\end{document}